\documentclass[twocolumn]{aastex701}
\usepackage{color}
\usepackage{graphicx}	
\usepackage{amsmath}	
\usepackage{amssymb}	
\usepackage{natbib}
\usepackage{enumerate}
\usepackage[version=4]{mhchem}
\usepackage{graphicx}
\usepackage{subcaption}

\begin{document}

\title{Centimeter-wave OH observations of comets 12P/Pons-Brooks and C/2023 A3 (Tsuchinshan-ATLAS) with FAST}

\author[0000-0002-9703-3110]{Long-Fei Chen}
\affiliation{School of Physics and Electronic Science, Guizhou Normal University, Guiyang 550025, China}
\affiliation{Guizhou Provincial Key Laboratory of Radio Astronomy and Data Processing, Guiyang 550025, China}
\email{chenlongfei@gznu.edu.cn}

\author[0009-0001-7333-5202]{Juncen Li}
\affiliation{Shanghai Astronomical Observatory, Chinese Academy of Sciences, 80 Nandan Road, Shanghai 200030, China}
\email{jcli@shao.ac.cn}

\author[]{Zhen Wang}
\affiliation{Key Laboratory of Radio Astronomy, Xinjiang Astronomical Observatory, Chinese Academy of Sciences, 150 Science 1-Street, Urumqi, Xinjiang 830011, China}
\email{wangzh@xao.ac.cn}

\author[0000-0003-3841-9977]{Jian-Yang Li}
\affiliation{Planetary Environmental and Astrobiological Research Laboratory (PEARL), School of Atmospheric Sciences, Sun Yat-sen University, Zhuhai 519082, China}
\affiliation{Key Laboratory of Radio Astronomy, Xinjiang Astronomical Observatory, Chinese Academy of Sciences, 150 Science 1-Street, Urumqi, Xinjiang 830011, China}
\email{lijianyang@mail.sysu.edu.cn}

\author[0000-0002-3140-5014]{Wing-Huen Ip}
\affiliation{Institute of Astronomy, National Central University, No. 300, Zhongda Road, Zhongli Dist., Taoyuan City 320317, Taiwan}
\email{wingip@astro.ncu.edu.tw}

\author[0000-0003-3827-8991]{Zhong-Yi Lin}
\affiliation{Institute of Astronomy, National Central University, No. 300, Zhongda Road, Zhongli Dist., Taoyuan City 320317, Taiwan}
\email{if207if@gmail.com}

\author[0000-0002-5033-9593]{Bin Yang}
\affiliation{Instituto de Estudios Astrof\'{i}sicos, Facultad de Ingenier\'{i}a y Ciencias, Universidad Diego Portales, Santiago, Chile}
\email{bin.yang@mail.udp.cl}

\correspondingauthor{Chao-Wei Tsai}
\author[0000-0002-9390-9672]{Chao-Wei Tsai}
\affiliation{State Key Laboratory of Radio Astronomy and Technology, National Astronomical Observatories, Chinese Academy of Sciences, Beijing 100101, China}
\affiliation{National Astronomical Observatories, Chinese Academy of Sciences, Beijing 100101, China}
\affiliation{Institute for Frontiers in Astronomy and Astrophysics, Beijing Normal University, Beijing 102206, China}
\email[show]{cwtsai@nao.cas.cn}


\begin{abstract}

We present centimeter-wave spectroscopic observations of the \ce{OH} 18-cm lines in two bright comets, 12P/Pons-Brooks and C/2023 A3 (Tsuchinshan-ATLAS), conducted with the Five-hundred-meter Aperture Spherical radio Telescope (FAST) during their 2024 apparitions.
For the Halley-type comet 12P/Pons-Brooks, five epochs of \ce{OH} observations were obtained. The main \ce{OH} lines at 1665 and 1667 MHz were robustly detected in absorption during one pre-perihelion epoch, while upper limits were derived for two post-perihelion epochs.
For the dynamically new Oort Cloud comet C/2023 A3 (Tsuchinshan-ATLAS), six epochs of \ce{OH} observations were obtained. The 1665 and 1667 MHz lines were robustly detected in absorption during two epochs immediately following the comet's closest approach to Earth. We also reported a tentative detection of the 1721 MHz satellite line in one of these two epochs.
The low \ce{OH} detection rates for these two bright comets are primarily attributable to their unfavorable heliocentric radial velocity during the observations, which resulted in the anti-maser negative inversion mode of the \ce{OH} excitation.
We calculated the \ce{OH} production rates for both comets and found that, after considering the small beam size of FAST and the collisional quenching effect, production rates are consistent with the data from literature. Specifically, this matches the power-law fit of the \ce{OH} production rates for comet 12P/Pons-Brooks, as well as the production rates at comparable heliocentric distances among comet C/2023 A3 (Tsuchinshan-ATLAS) and four other dynamically new Oort Cloud comets.

\end{abstract}



\section{Introduction}\label{intro}
Our Solar System has evolved for billions of years, since its birth from the Solar Nebula to the proto-star, and from the proto-planetary disk to our present planetary system \citep{Ehrenfreund2000, Chambers2023, Biver2024a}. Comets are icy bodies with dust grains mixed with water ices, and volatile and refractory species that represent the oldest and most primitive materials since their formation \citep{Altwegg2022}, and have not undergone significant reprocessing due to their cold formation and storage environments \citep{Wang2017, Biver2024a}. Following these evolutionary footprints, comets as the target vessels serve as `time capsules' that preserve the information of the early formation conditions of our Solar System to the material compositions of present rocky planets. Consequently, comets can be transferred to the inner solar system by gravitational scattering. Therefore, comets may play an important role in the delivery of water and organic material to the early Earth \citep{Morbidelli2000, Mandt2024}. The isotopic ratios of molecules found in comet materials may also provide clues that trace the evolutionary history of our Solar System \citep{Bockelee-Morvan2015}.

When approaching the Sun, icy grains can sublimate from the surface of comets due to solar heating, leading to the formation of the cometary coma. In the sublimated materials, water (\ce{H2O}) is one of the most abundant species in these icy grains. It can be further photon-dissociated into atomic hydrogen (\ce{H}) and hydroxyl radical (\ce{OH}), or \ce{H2} and \ce{O} \citep{Schleicher1988, Huebner1992}. Among these species, only \ce{H} and \ce{OH} can be observed by radio telescopes at centimeter wavelengths. However, \ce{H} emission in the cometary coma can be contaminated by the similar emission from the background interstellar medium, leaving \ce{OH} as the only species that can be used as a proxy for water production in comets. \ce{OH} has four hyperfine transitions (1612.231, 1665.402, 1667.359, and 1720.530 MHz) at the 18 cm radio wavelength, which provide useful information on the velocity and distribution of the cometary coma. Under local thermodynamic equilibrium (LTE) conditions, the line strengths of these four transitions follow a ratio of 1:5:9:1 \citep{Tang2021}. Therefore, the \ce{OH} transitions at 1665.402 and 1667.359 MHz are the main lines that have been widely detected in cometary coma since their first detection in comet C/1973E1 Kohoutek \citep{Biraud1974}.

During 2024, we monitored two comets, 12P/Pons-Brooks and C/2023 A3 (Tsuchinshan-ATLAS), using the Five-hundred-meter Aperture Spherical radio Telescope (FAST). Both comets were naked-eye-visible in 2024. While 12P/Pons-Brooks exhibited frequent outbursts and abundant molecular emission, C/2023 A3 (Tsuchinshan-ATLAS) displayed nearly steady-state dust emission before and after perihelion \citep{Liu2025, Moreno2025} under carbon-depleted conditions \citep{Ahuja2024, Tang2024, Cambianica2025}. Detailed summaries of both are provided below.

Comet 12P/Pons-Brooks (hereafter referred to as 12P) is one of the brightest comets in the first half of 2024, with a peak visual magnitude of 3.5 mag on April 6, 2024\footnote{https://astro.vanbuitenen.nl/comet/12}. It is a Halley-type comet with a period of 71.2 years and reaches perihelion on April 21, 2024. Its estimated rotation period is approximately 57 hours \citep{Knight2024}. Before and after its perihelion, it exhibited multiple large and minor outbursts \citep{Ferellec2024, Gritsevich2025, Jewitt2025}. Several species have been reported in the optical wavelength \citep{Jehin2023a, Jehin2023b, Jehin2023c, Jehin2024a, Jehin2024b, Ferellec2024, Zhao2025, VanderDonckt2026}, including \ce{OH}, \ce{C2}, \ce{C3}, \ce{CN}, and \ce{NH2}. Dozens of species at millimeter and centimeter wavelengths were also reported by \citet{Biver2024b} and \citet{Li2025}, including \ce{HCN}, \ce{HNC}, \ce{HC3N}, \ce{CH3CN}, \ce{HNCO}, \ce{NH2CHO}, \ce{CH3OH}, \ce{H2CO}, \ce{CO}, \ce{HCO+}, \ce{CS}, \ce{H2S}, \ce{OCS}, \ce{SO}, \ce{C^34S}, \ce{OH}, and \ce{NH3}.

On the other hand, the comet C/2023 A3 (Tsuchinshan-ATLAS) (hereafter referred to as A3) is the brightest comet in the second half of 2024, with a peak visual magnitude of -4.9 mag on October 9, 2024. This peak occurred after its perihelion on September 27, 2024 and before its nearest approach to Earth on October 12, 2024\footnote{https://astro.vanbuitenen.nl/comet/2023a3}. This comet is a dynamically new comet with Oort Cloud origin \citep{Munaretto2026}, and is evolving into a hyperbolic orbit based on the NASA/JPL ephemeris. Because of its great brightness, this comet has attracted much attention from the community. Based on its anomalously bright light curve before perihelion, early studies suggested that it was experiencing fragmentation and could disintegrate \citep{Sekanina2024a, Sekanina2024b}. However, with additional constrains, \citet{Liu2025} showed that the nucleus likely remained stable, which is consistent with follow-up observations on this comet \citep{Moreno2025}. The optical spectroscopic observations showed a variation on the molecular production rate for this comet. \citet{Jehin2024c} reported the detections of \ce{OH}, \ce{CN}, and \ce{C2} at a heliocentric distance of $r_h = 1.81$ AU before its perihelion. In addition, the observations also showed that this comet was in a carbon-depleted situation, with only \ce{CN} and \ce{Na} lines detected, while \ce{C2}, \ce{C3}, and \ce{NH2} were only upper limit constrained \citep{Ahuja2024, Tang2024, Cambianica2025}. \citet{Ahuja2024} calculated a water production rate of $1.5 \times 10^{28}$ molecules s$^{-1}$ at $r_h = 2.33$ AU during this period, which is consistent with the value inferred from OH production rate by \citet{Jehin2024c}. Despite the paucity of pre-perihelion molecular detections, over a dozen species were identified post-perihelion. After the nearest approach, there were reports that \ce{O}, \ce{C2}, and \ce{NH2} were detected in the optical spectra \citep{Mugrauer2024a, Mugrauer2024b}. \citet{Munaretto2026} reported the identification of Na I, [OI], \ce{CN}, \ce{C2}, \ce{C3} and \ce{NH2} at $r_h =0.83$ AU. \citet{Kobayashi2025} also reported the detection of \ce{H2O}, \ce{OH}, \ce{HCN}, \ce{C2H2}, \ce{CH4}, \ce{C2H6}, and \ce{CH3OH} in near-infrared wavelength band at a $r_h = 0.91$ AU. To date, there have been no radio observational reports for this comet.

In this paper, we report the results of centimeter wavelength observations of \ce{OH} for comets 12P and A3 using FAST. Section \ref{sec.obs} presents the observational schedules. The data processing and results are reported in Section \ref{sec.dat} and \ref{sec.res}, respectively, followed by a discussion in Section \ref{sec.disc} and a summary in Section \ref{sec.sum}.

\section{Observations}\label{sec.obs}
FAST is a radio telescope similar in design to Aricebo, but with larger diameter, higher sensitivity, and more sky coverage. This project was an early application of the non-sidereal target tracking system \citep{Chen2024} and therefore observation times were not chosen to optimize the excitation conditions of the OH lines (see Section \ref{subsec.exc} for details on the impact of the results).

\begin{figure*}
\centering
\includegraphics[scale=0.3]{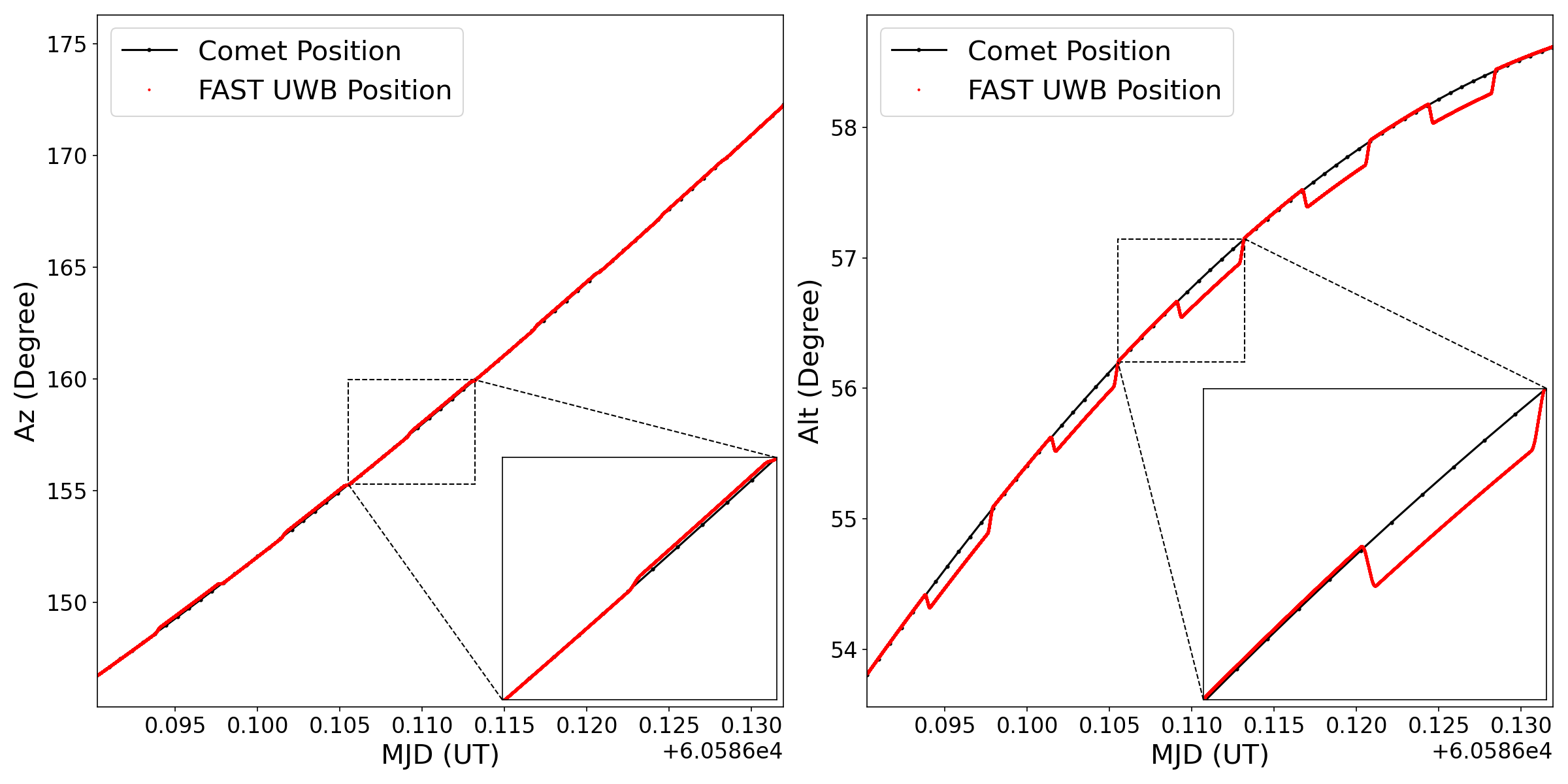}
\caption{The telescope tracking (red) and the local Az and Alt coordinates (black) for comet C/2023 A3 (Tsuchinshan-ATLAS) on October 3, 2024. The sub-figure in each panel shows the zoom in of a cycle of source ON and OFF.
}
\label{tracking_A3}
\end{figure*}

We observed comets 12P and A3 in April and May, and October of 2024 with FAST, respectively. The ultra-wide bandwidth (UWB) receiver \citep{Zhang2023} was used as the backend during the observations. This receiver simultaneously covers 500--3300 MHz, allowing for the possibility of \ce{OH} detection. Observations of comet 12P were performed over a total of five days, with each day consisting of $\sim$1 hour of tracking. For comet A3, observations spanned six days. Table \ref{log} summarizes the observation logs for comets 12P and A3. The ephemeris for each comet were obtained from the HORIZONS system of Jet Propulsion Laboratory\footnote{https://ssd.jpl.nasa.gov/horizons/app.html}. To track the moving comets, a custom observation mode was used to improve accuracy and efficiency. Detailed descriptions of this mode can be found in \citet{Chen2024}. In brief, we pre-designed the tracking positions of the observed comets, namely the local Alt and Az coordinates, based on their ephemeris before each observation. To obtain background continuum subtraction, we performed five minutes of source ON and five minutes of source OFF. Between the source ON and OFF integration times, a 30 second switching time was used to separate the source ON and OFF positions by 10$^\prime$.22 to avoid the exceeding of the acceleration limit of 90$^{\prime\prime}$ per second for the FAST receiver maneuver.
Assuming an \ce{OH} scale-length of 1.4$\times$10$^{5}$ km \citep{Bockelee-Morvan1990}, this corresponds to an angular size of 2.57$^\prime$ and 4.63$^\prime$ for the comet 12P and A3 at the observations, respectively. Therefore, the setup of the angular separation between the source ON and OFF is acceptable. We also checked the spectra from the source OFF position to ensure that no emission or absorption were present. Figure \ref{tracking_A3} shows the tracking positions for comet A3 on October 3, 2024.

\begin{table*}
\caption[]{Observation logs for comets 12P/Pons-Brooks and C/2023 A3 (Tsuchinshan-ATLAS) using FAST.}
\label{log}
\begin{tabular}{lllll}
\hline\noalign{\smallskip}
Start observing  & Duration  & Heliocentric   & Geocentric    & $\dot{\Delta}$$^a$\\
UT time          & (hours)   & distance (AU)  & distance (AU) & (km s$^{-1}$)\\
\hline\noalign{\smallskip}
\multicolumn{5}{l}{Comet 12P/Pons-Brooks} \\
\hline\noalign{\smallskip}
2024-04-17 06:53 & 1.00      & 0.78           & 1.61          & -0.73 \\
2024-04-24 07:22 & 1.00      & 0.78           & 1.60          & -1.58 \\
2024-05-10 05:00 & 1.00      & 0.86           & 1.58          & -3.51 \\
2024-05-11 05:00 & 1.00      & 0.87           & 1.58          & -3.54 \\
2024-05-13 05:00 & 0.80      & 0.88           & 1.57          & -3.54 \\
\hline\noalign{\smallskip}
\multicolumn{5}{l}{Comet C/2023 A3 (Tsuchinshan-ATLAS)} \\
\hline\noalign{\smallskip}
2024-10-03 02:10 & 1.00      & 0.42           & 0.68          & -66.85 \\
2024-10-04 02:10 & 1.00      & 0.43           & 0.65          & -62.95 \\
2024-10-13 05:00 & 1.43      & 0.57           & 0.47          & 4.97   \\
2024-10-14 05:52 & 1.43      & 0.59           & 0.48          & 13.39  \\
2024-10-16 06:35 & 1.52      & 0.62           & 0.50          & 27.94  \\
2024-10-17 06:29 & 1.33      & 0.64           & 0.52          & 33.96  \\
\hline\noalign{\smallskip}
\end{tabular}\\
Notes: $^a$ Line-of-sight velocity of the target with respect to the observer. A positive value means the target center is moving away from the observer, negative indicates movement toward the observer. Data were taken from JPL/Horizons ephemeris.
\end{table*}

\section{Data processing}\label{sec.dat}
\begin{figure*}
\centering
\begin{subfigure}{0.45\textwidth}
\includegraphics[width=\textwidth]{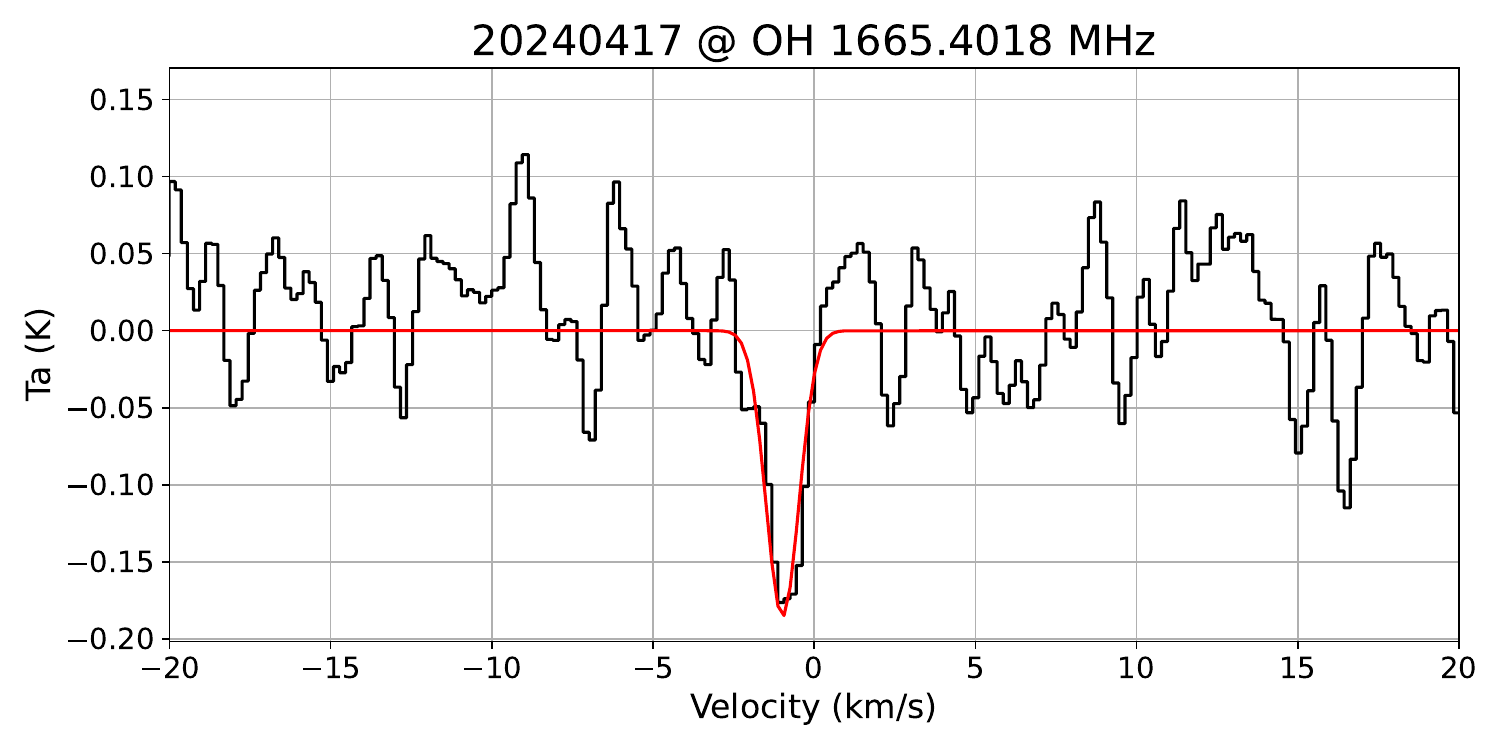}
\end{subfigure}
\begin{subfigure}{0.45\textwidth}
\includegraphics[width=\textwidth]{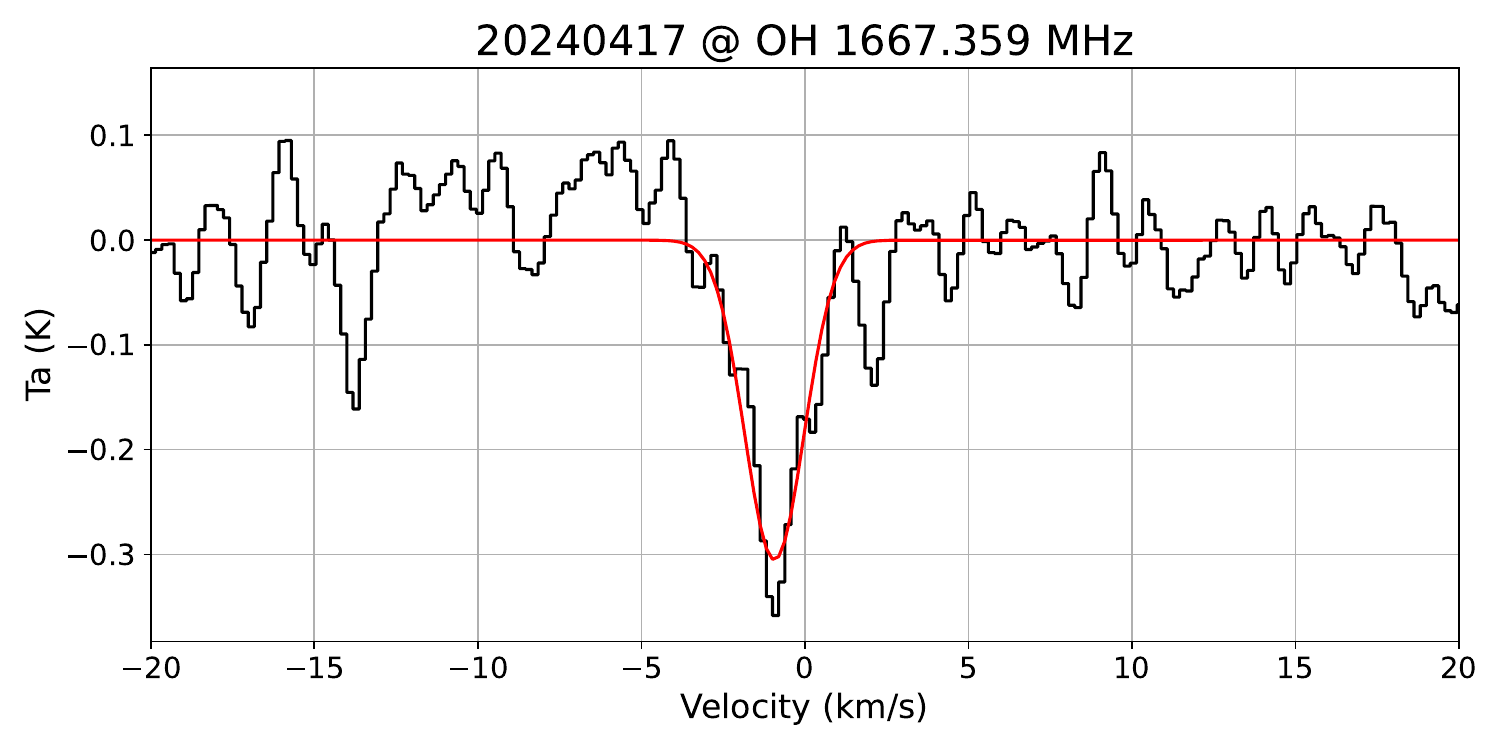}
\end{subfigure}

\begin{subfigure}{0.45\textwidth}
\includegraphics[width=\textwidth]{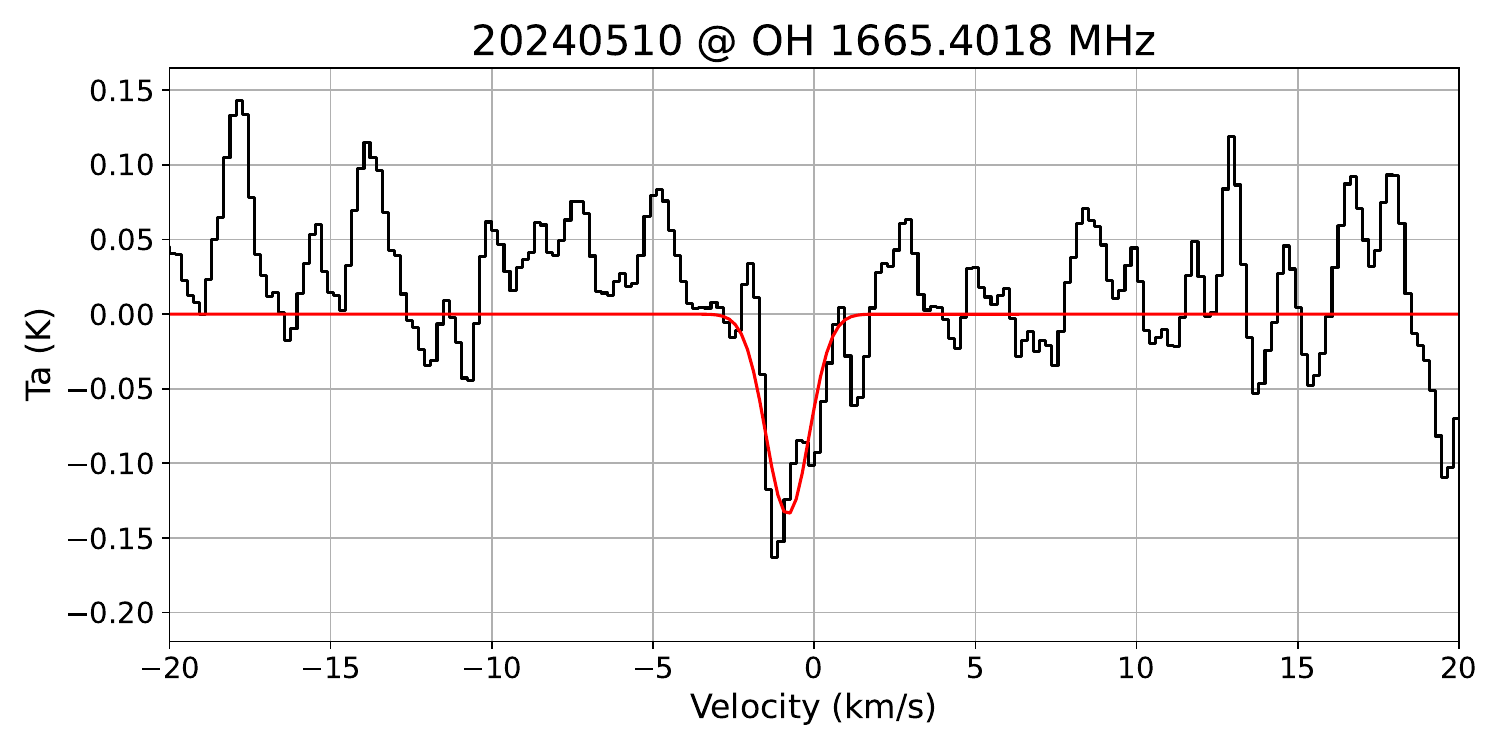}
\end{subfigure}
\begin{subfigure}{0.45\textwidth}
\includegraphics[width=\textwidth]{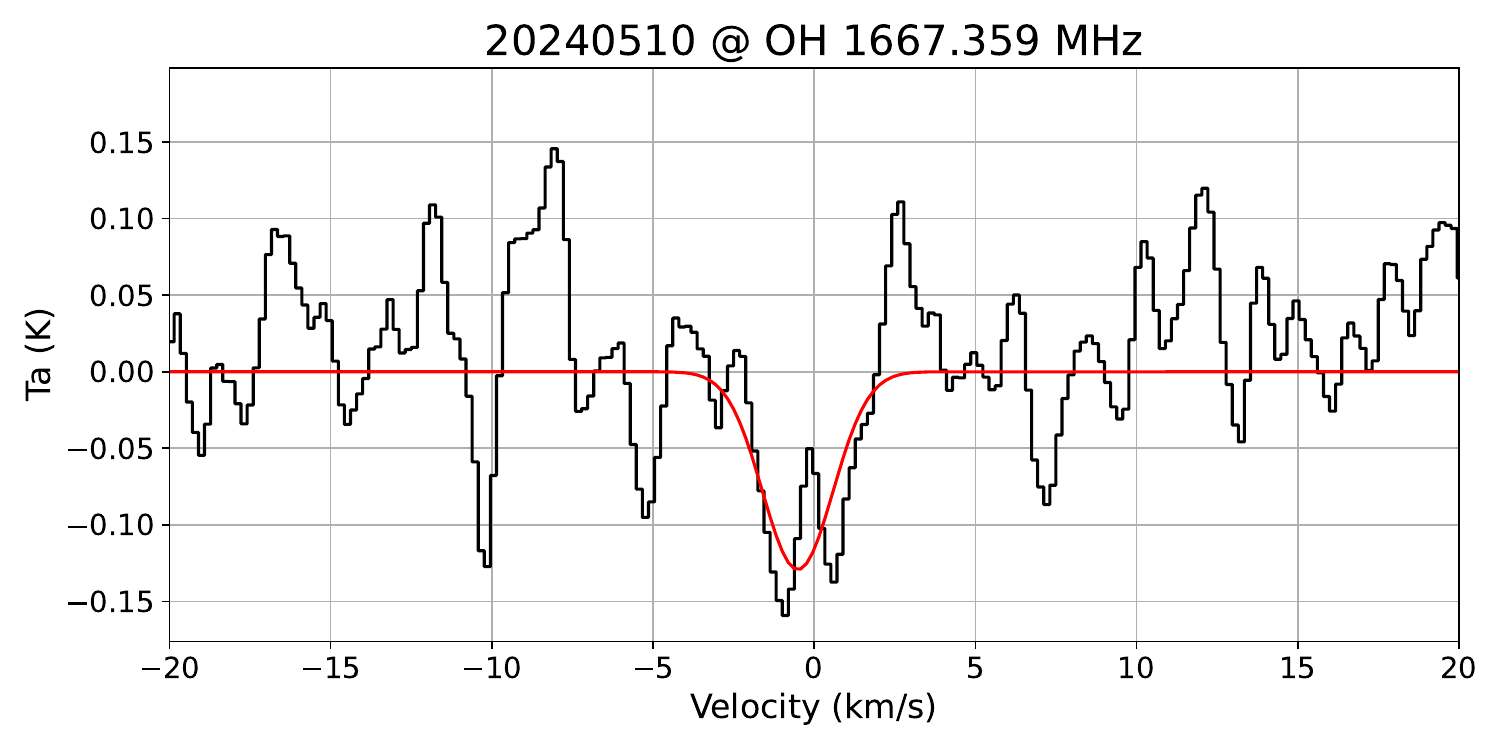}
\end{subfigure}

\begin{subfigure}{0.45\textwidth}
\includegraphics[width=\textwidth]{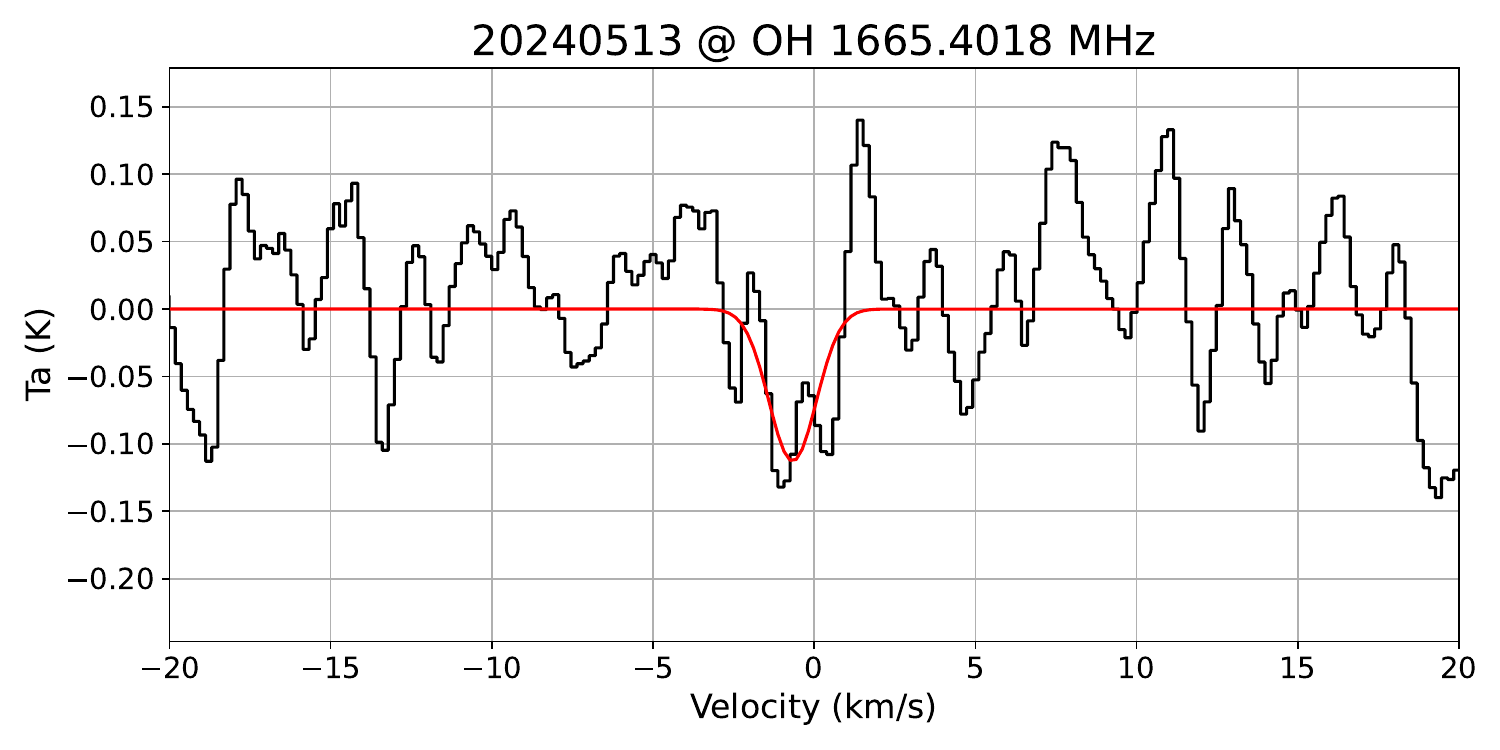}
\end{subfigure}
\begin{subfigure}{0.45\textwidth}
\includegraphics[width=\textwidth]{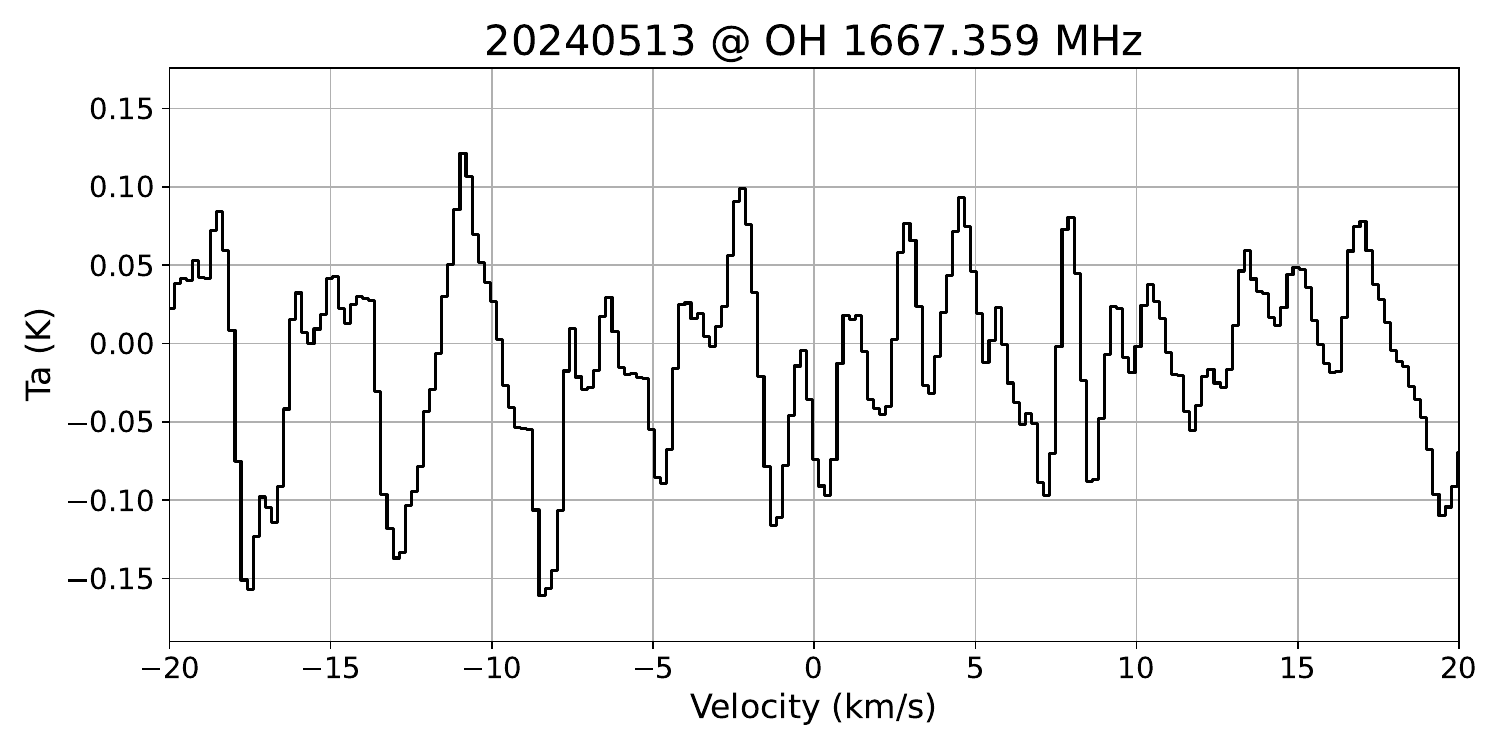}
\end{subfigure}

\caption{The \ce{OH} absorption lines with detection, tentative detection, and non-detection for comet 12P/Pons-Brooks. The black line shows the smoothed spectra, while the red line is the Gaussian fitting.}
\label{OH_12P}
\end{figure*}

\begin{figure*}
\centering
\begin{subfigure}{0.45\textwidth}
\includegraphics[width=\textwidth]{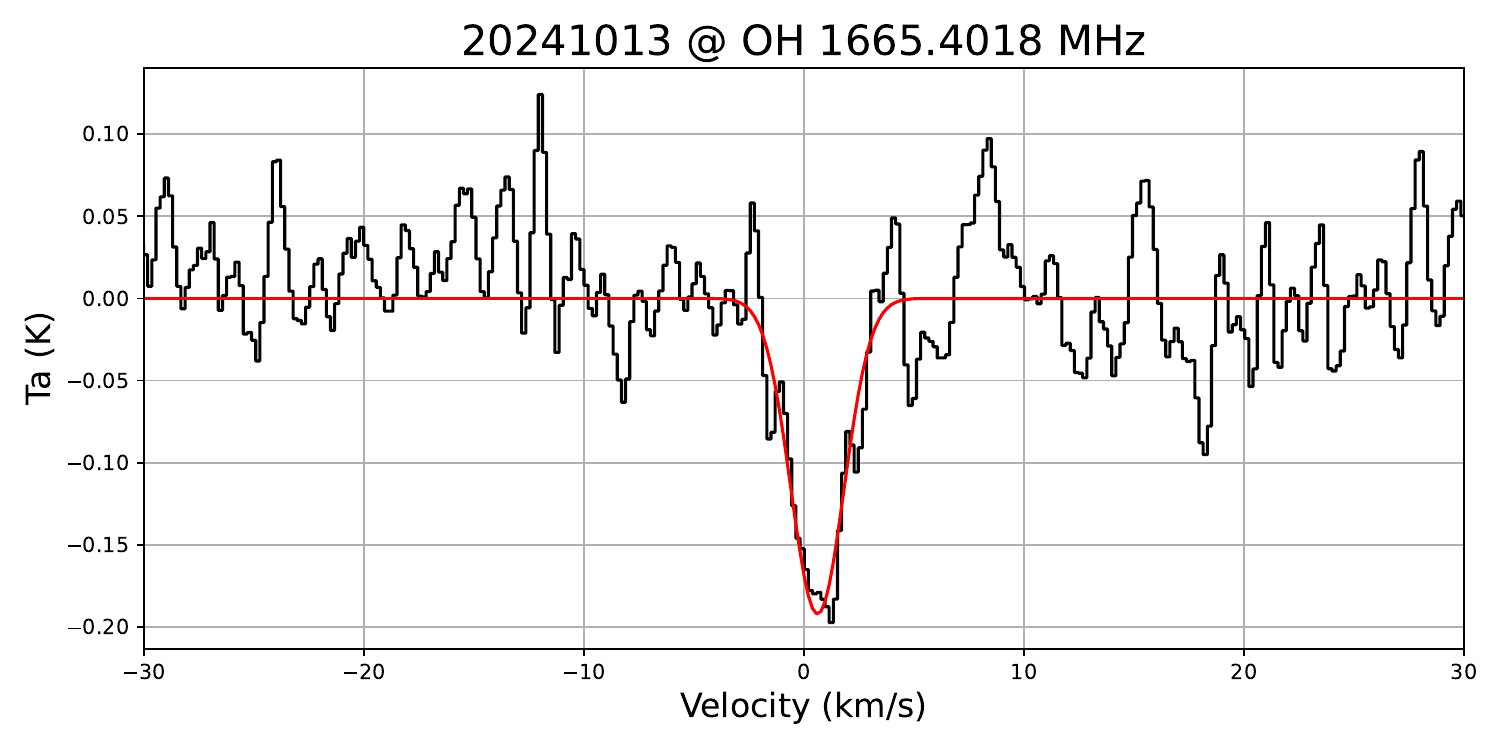}
\end{subfigure}
\begin{subfigure}{0.45\textwidth}
\includegraphics[width=\textwidth]{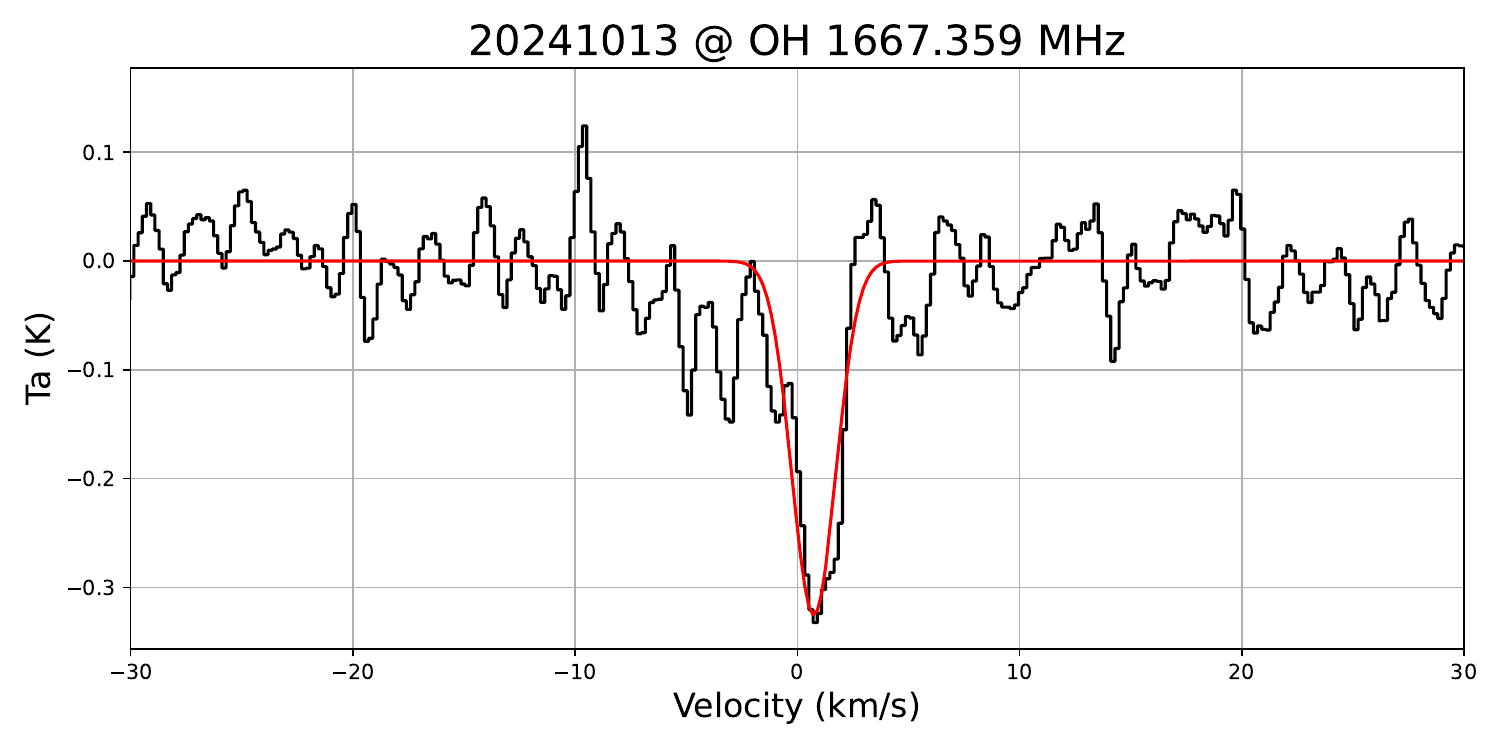}
\end{subfigure}

\begin{subfigure}{0.45\textwidth}
\includegraphics[width=\textwidth]{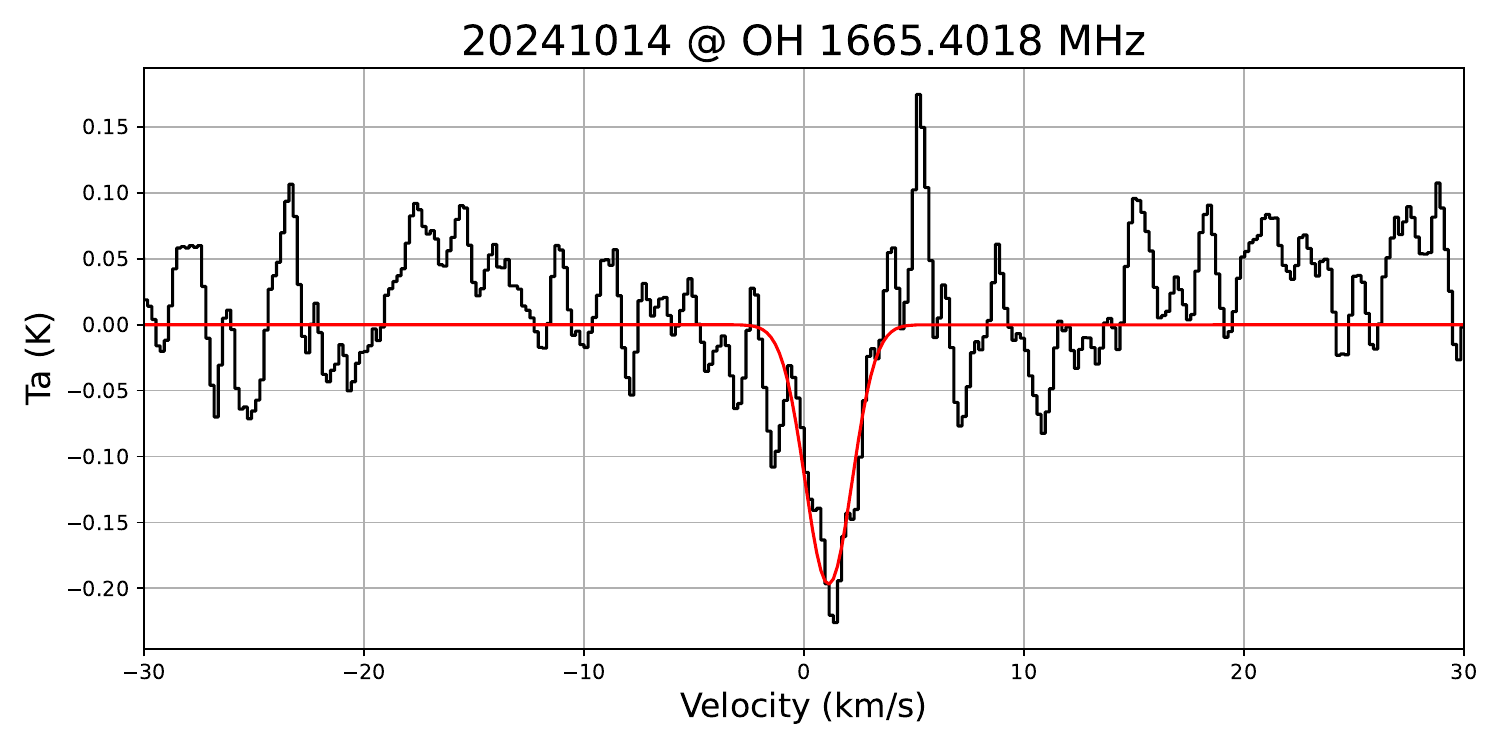}
\end{subfigure}
\begin{subfigure}{0.45\textwidth}
\includegraphics[width=\textwidth]{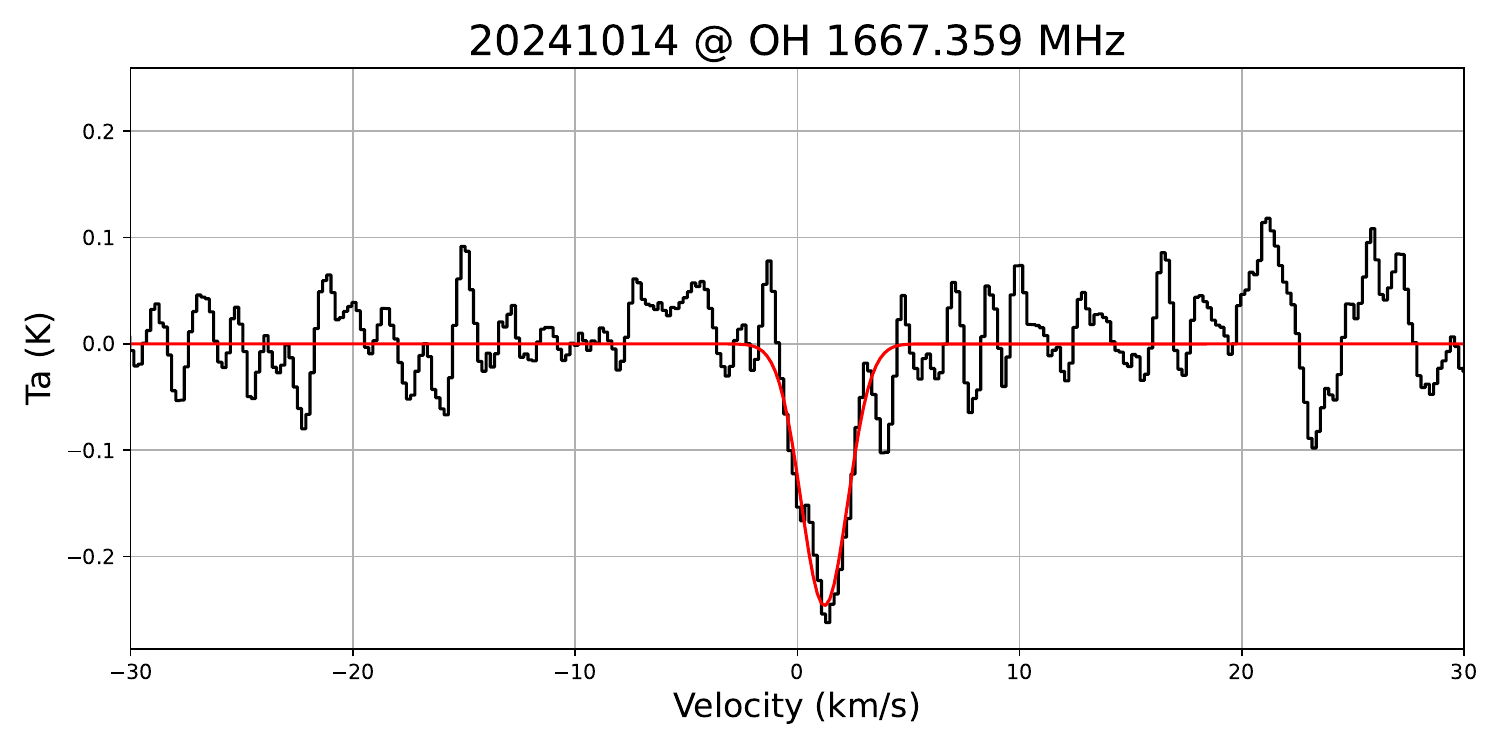}
\end{subfigure}

\caption{The \ce{OH} absorption lines with detection for comet C/2023 A3 (Tsuchinshan-ATLAS). The black line shows the smoothed spectra, while the red line is the Gaussian fitting.}
\label{OH_A3}
\end{figure*}

We followed the data processing pipeline presented in \citet{Chen2024} except that the two polarizations (XX and YY) were separately calibrated before being combined into the final spectra. The raw data were first calibrated to antenna temperature, followed by Doppler velocity correction with respect to the comet nucleus. The continuum was subtracted based on the source OFF positions. The antenna temperature could be further converted to flux density using the antenna gain described in \citet{Zhang2023}, which is essential for calculating the \ce{OH} molecular production rate. Finally, the final spectra were boxcar-smoothed with 4 bins, corresponding to a velocity resolution of 0.76 km s$^{-1}$. In this paper, we focused on the \ce{OH} transitions. We examined the observational data from all the days for the two comets, 12P and A3. However, the \ce{OH} spectral lines were only detected or tentatively detected on three out of five days for comet 12P and two out of six days for comet A3. Figure \ref{OH_12P} and Figure \ref{OH_A3} show the \ce{OH} spectral lines at 1665.402 and 1667.359 MHz for comets 12P and A3, respectively. Table \ref{line_profile} presents the fitted line profile parameters for these two comets.

For comet 12P, there was only one observation day before its perihelion, on which the two main \ce{OH} lines were successfully detected (April 17, 2024). For the four observation days after the perihelion, tentative detections were made on two days, specifically on May 10 and May 13, 2024. Although the signal-to-noise ratio (SNR) of the spectra on May 10 is 2.7 for 1667.359 MHz transition line, the center velocity and line width are generally consistent with the detections on April 17. On May 13, only the \ce{OH} line at 1665.402 MHz was tentatively detected, with an SNR of 2.0, while no signal was identified for the \ce{OH} line at 1667.359 MHz on this day.

For comet A3, all six observation days were scheduled after its perihelion, with successful detections of the two main \ce{OH} lines occurring on two days immediately following its closest approach to Earth, namely on October 13 and October 14, 2024. These detections were characterized by high SNR.

\begin{figure*}
\centering
\begin{subfigure}{0.45\textwidth}
\includegraphics[width=\textwidth]{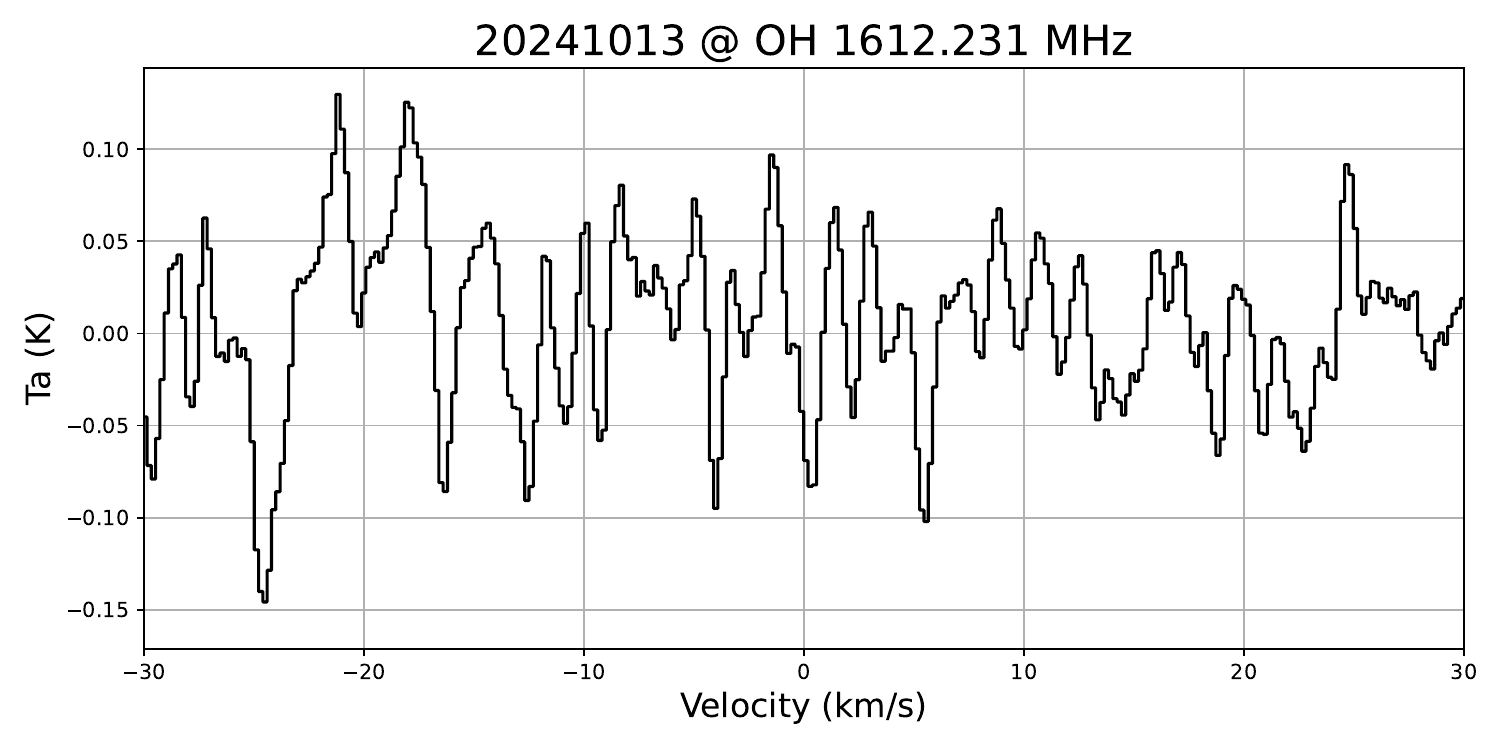}
\end{subfigure}
\begin{subfigure}{0.45\textwidth}
\includegraphics[width=\textwidth]{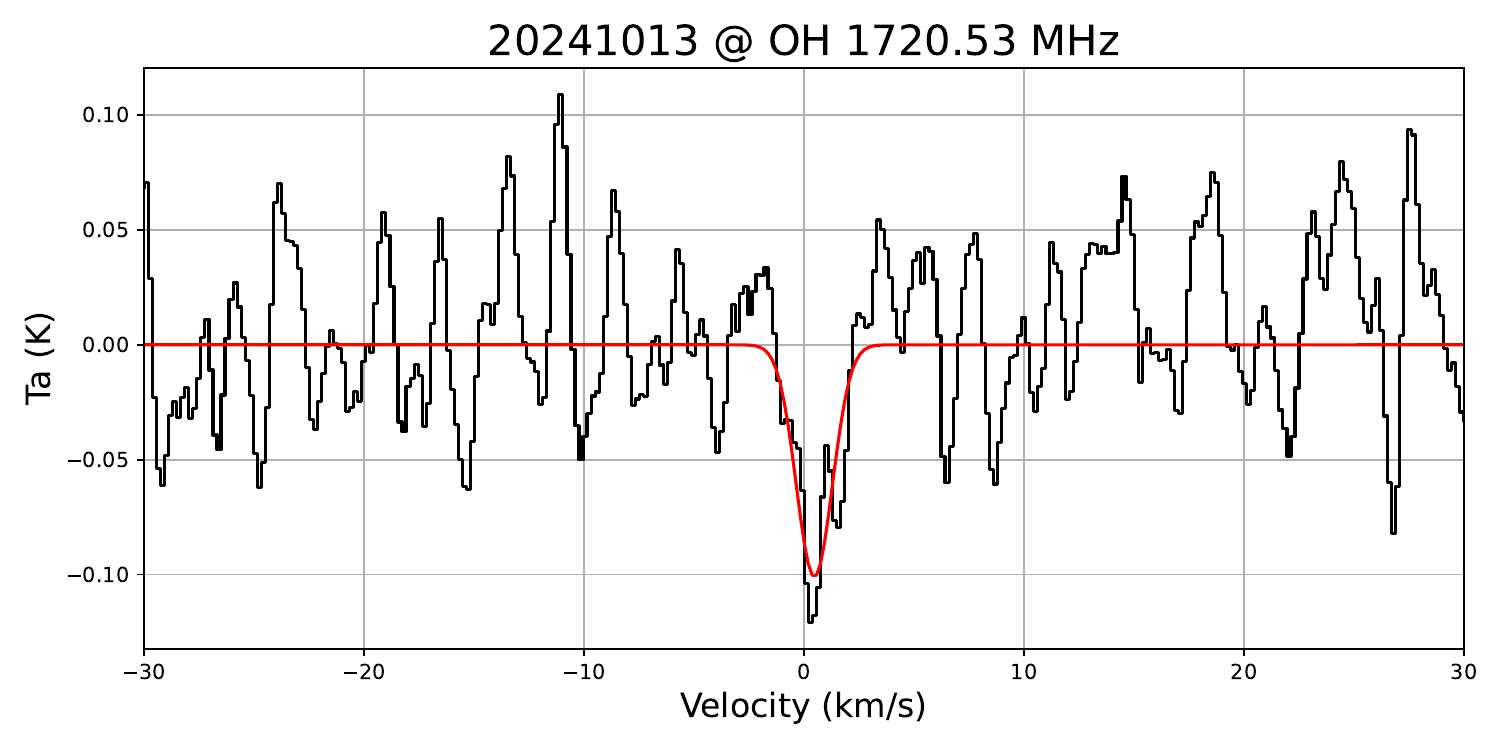}
\end{subfigure}
\caption{The \ce{OH} spectra for the 1612 and 1721 MHz lines for comet C/2023 A3 (Tsuchinshan-ATLAS).
The black line shows the smoothed spectra, while the red line is the Gaussian fitting.}
\label{OH_A3_1721}
\end{figure*}

\begin{figure*}
\centering
\begin{subfigure}{0.45\textwidth}
\includegraphics[width=\textwidth]{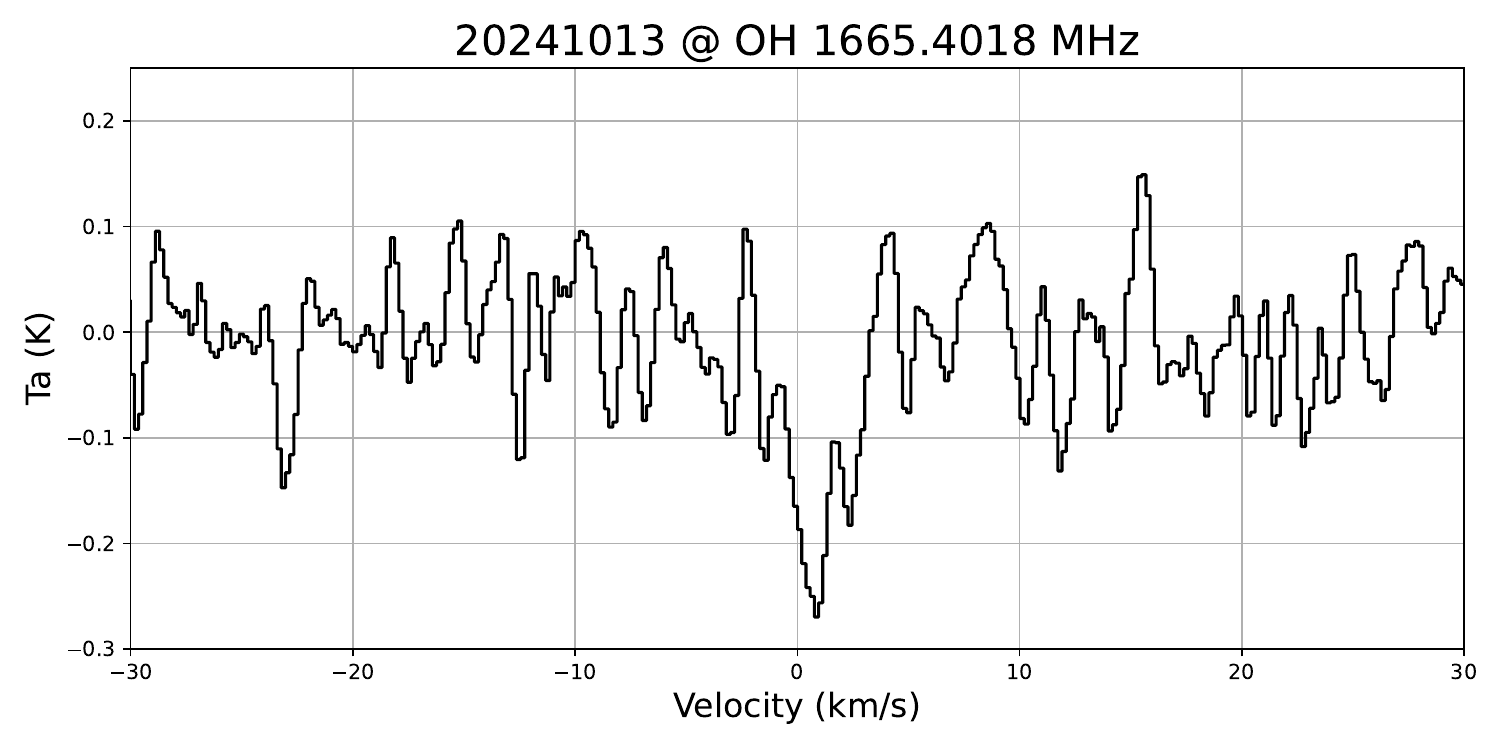}
\end{subfigure}
\begin{subfigure}{0.45\textwidth}
\includegraphics[width=\textwidth]{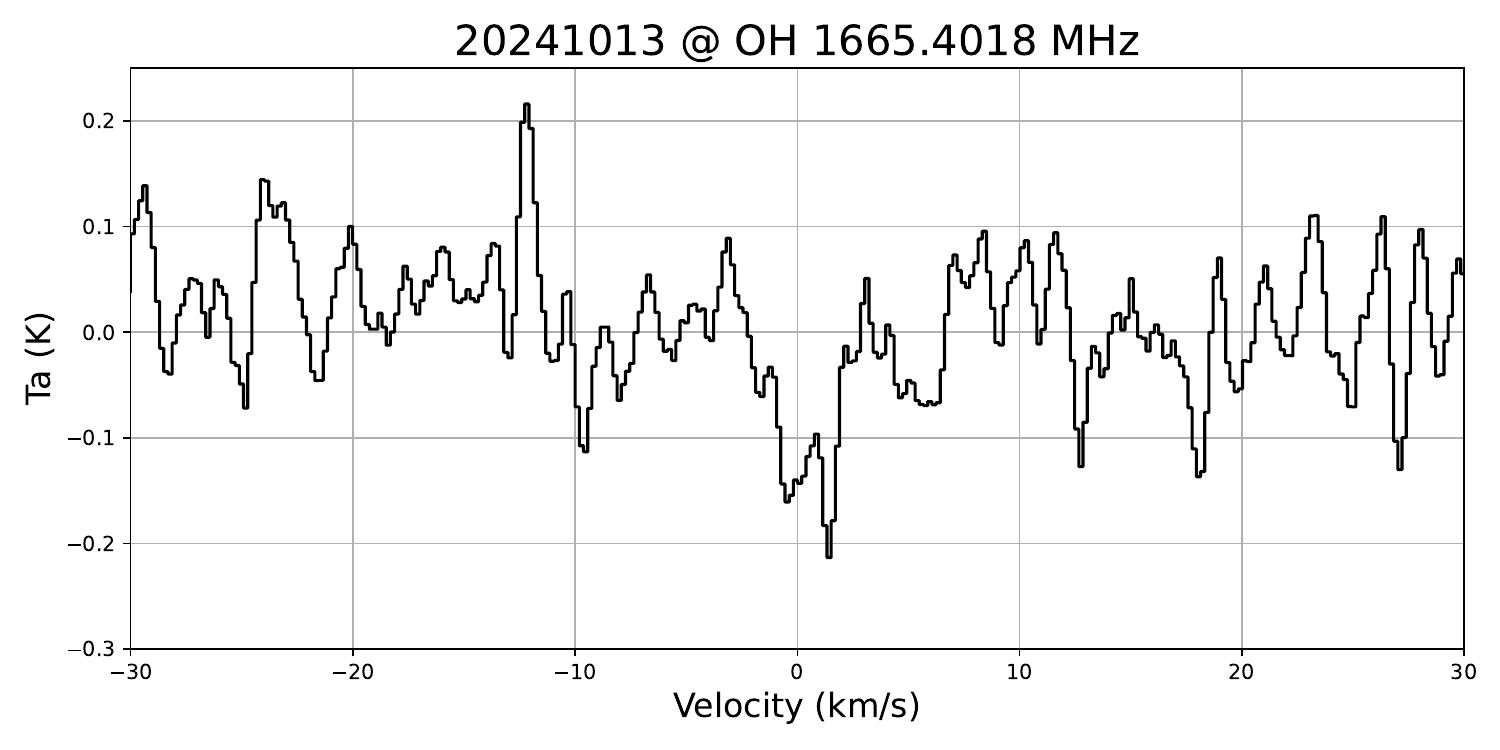}
\end{subfigure}

\begin{subfigure}{0.45\textwidth}
\includegraphics[width=\textwidth]{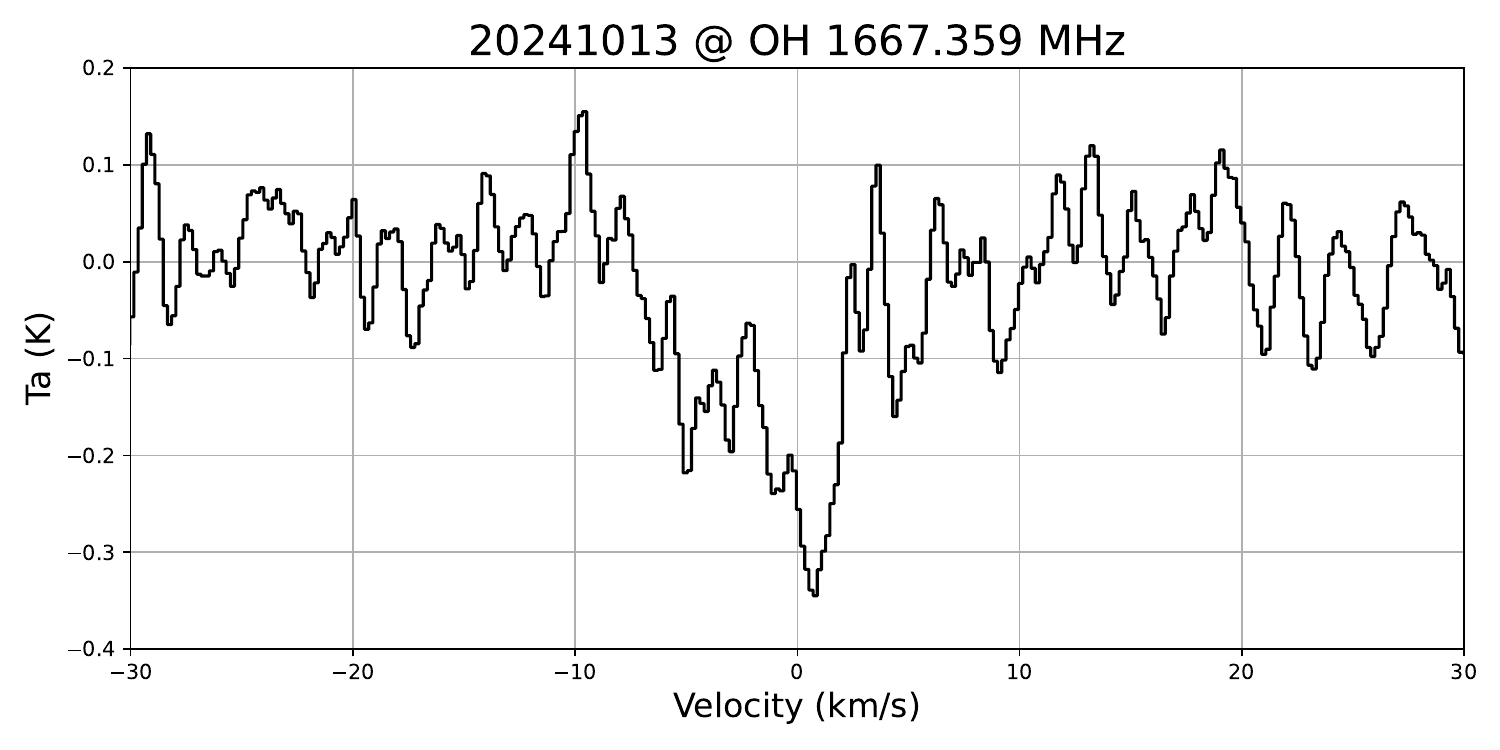}
\end{subfigure}
\begin{subfigure}{0.45\textwidth}
\includegraphics[width=\textwidth]{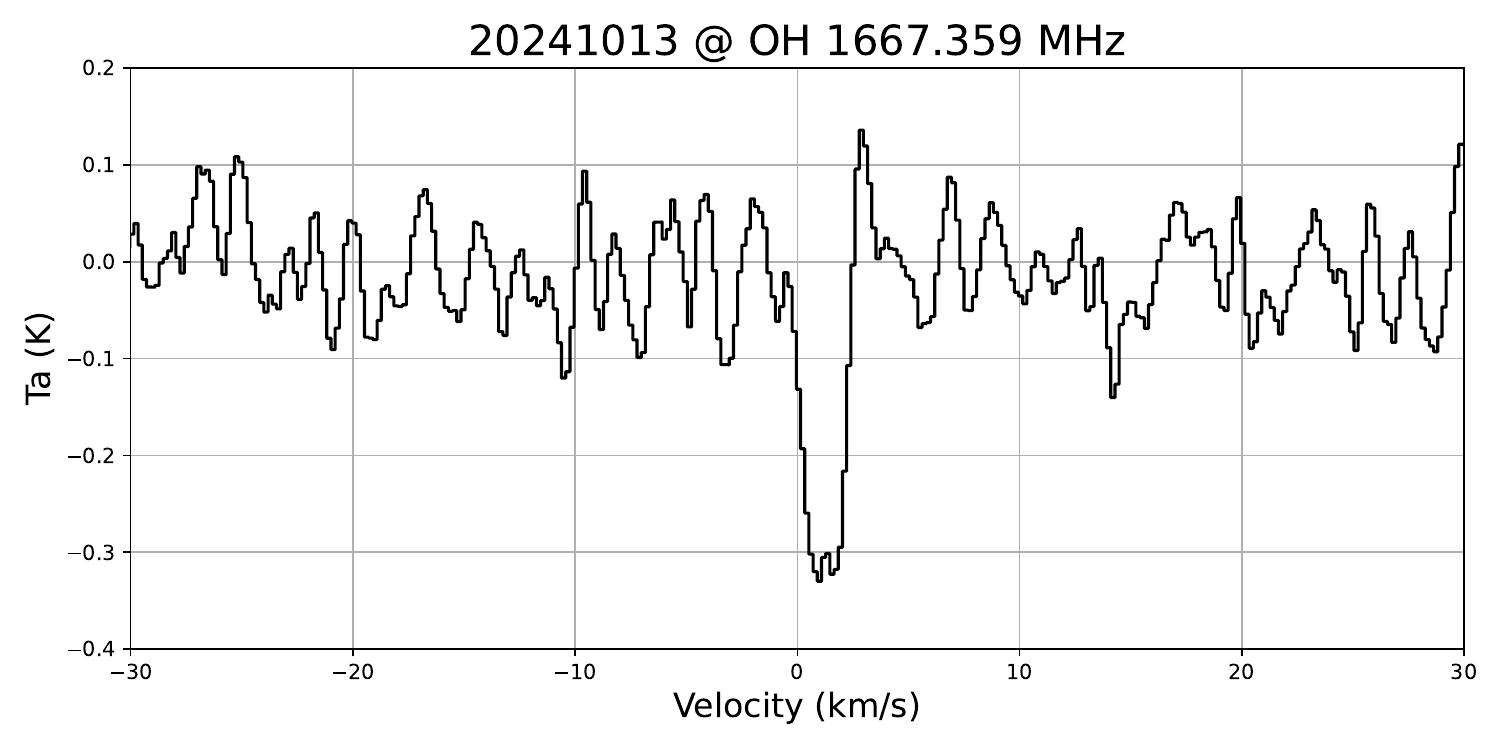}
\end{subfigure}

\begin{subfigure}{0.45\textwidth}
\includegraphics[width=\textwidth]{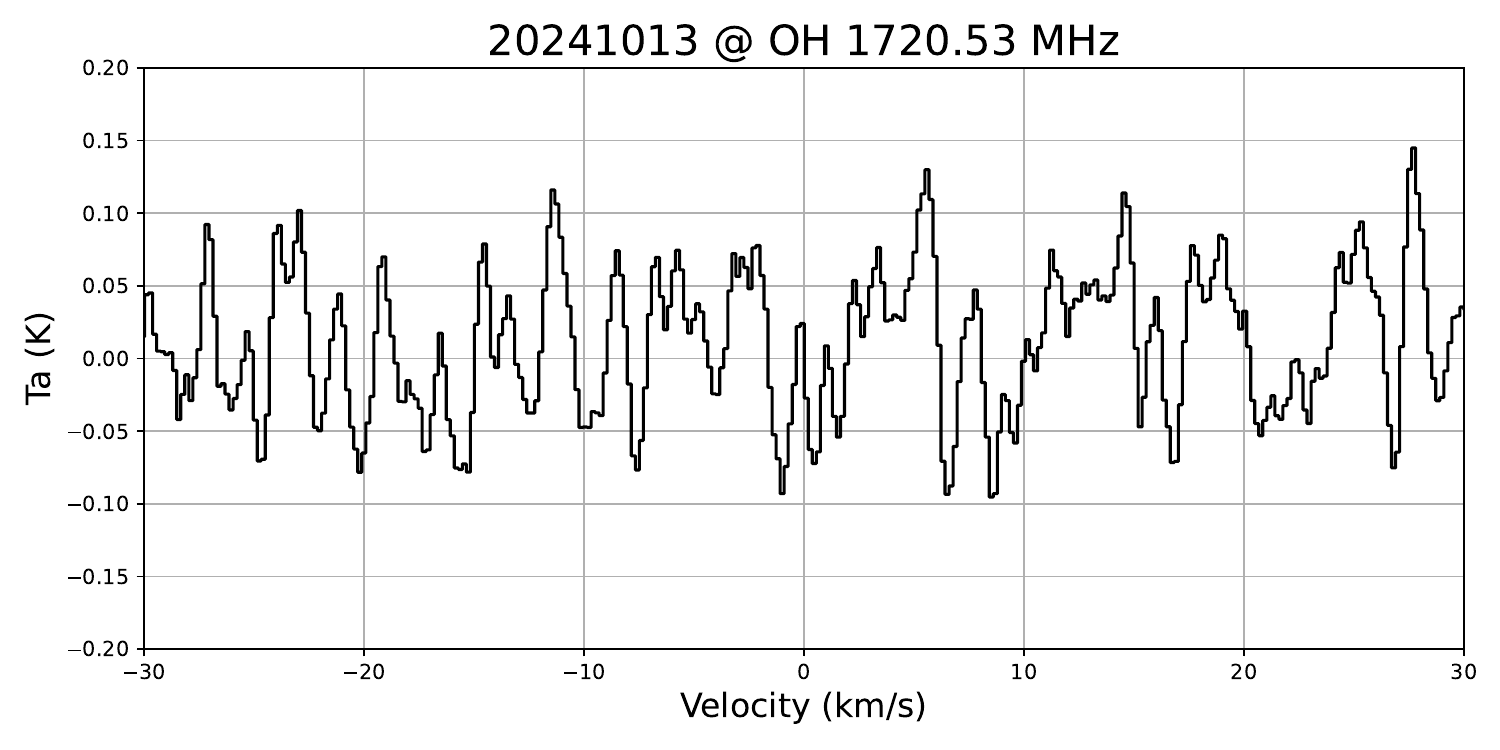}
\end{subfigure}
\begin{subfigure}{0.45\textwidth}
\includegraphics[width=\textwidth]{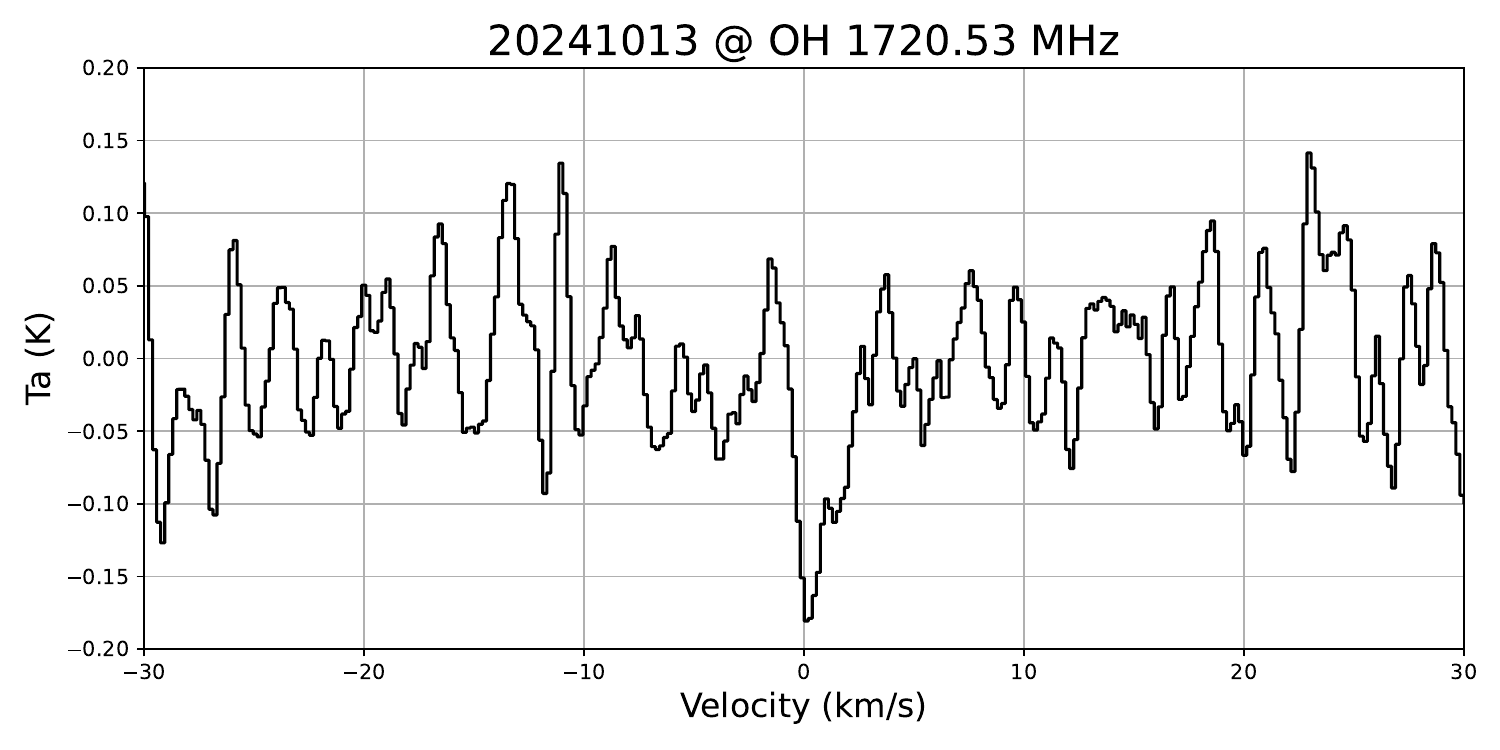}
\end{subfigure}
\caption{The XX and YY polarizations of the \ce{OH} 1665, 1667, and 1721 MHz line for comet C/2023 A3 (Tsuchinshan-ATLAS) on October 13, 2024. The left and right columns correspond to the XX and YY polarizations, respectively. The blueshifted broad line wing in the XX polarization of the 1667 MHz line is caused by the noise fluctuations of the spectra at the source OFF position.}
\label{OH_A3_1721_pola}
\end{figure*}

We also examined the two \ce{OH} satellite lines at 1612 and 1721 MHz for the two comets. No detections of these lines for comet 12P during the observation days. For comet A3, however, there was a tentative detection at 1721 MHz with a SNR of 2.7 on May 13, 2024. Figure \ref{OH_A3_1721} shows the spectra, and the Gaussian-fitted line profile parameters are shown in Table \ref{line_profile}. To further verify its detection, we checked the XX and YY polarizations of this line as showed in Fig.~\ref{OH_A3_1721_pola}, along with the comparison of the two polarizations for the 1665 MHz and the 1667 MHz lines. We found that the absorption was only present in the YY polarization for the 1721 MHz line. The fitted line profiles are slightly different than that for the 1667 MHz line. We will discuss the \ce{OH} excitation and the 1721 MHz line in Section \ref{subsec.exc}.

\begin{table*}
\caption[]{The \ce{OH} line profile parameters for comets 12P/Pons-Brooks and C/2023 A3 (Tsuchinshan-ATLAS) with detections.}
\label{line_profile}
\begin{tabular}{lllllllll}
\hline\noalign{\smallskip}
Obs. date  & \ce{OH} line  & $\nu$       & FWHM         & Peak Ta  & $\sigma$ & SNR  & S$_{int}$  & Q\\
           & (MHz)         & (km s$^{-1}$) & (km s$^{-1}$)  & (mK)     & (mK)     &      & (mJy km s$^{-1}$) & ($10^{28}$ molecules s$^{-1}$)\\
\hline\noalign{\smallskip}
\multicolumn{9}{l}{Comet 12P/Pons-Brooks} \\
\hline\noalign{\smallskip}
2024-04-17 & 1665.402  & -0.98$\pm$0.08  & 1.20$\pm$0.19  & -185.63$\pm$25.46  & 46.36  & 4.0 & -19.3$\pm$1.9 & -- \\
           & 1667.359  & -0.93$\pm$0.06  & 2.14$\pm$0.15  & -305.33$\pm$18.57  & 48.88  & 6.3 & -56.8$\pm$2.6 & 41.2$\pm$1.9 \\
\hline\noalign{\smallskip}
2024-05-10 & 1665.402  & -0.82$\pm$0.14  & 1.57$\pm$0.32  & -134.23$\pm$23.69  & 48.87  & 2.8 & -18.4$\pm$2.3 & -- \\
           & 1667.359  & -0.51$\pm$0.18  & 2.57$\pm$0.42  & -129.22$\pm$18.12  & 48.37  & 2.7 & -28.8$\pm$2.8 & 10.9$\pm$1.1 \\
\hline\noalign{\smallskip}
2024-05-13 & 1665.402  & -0.68$\pm$0.20  & 1.73$\pm$0.46  & -112.84$\pm$26.01  & 56.16  & 2.0 & -17.0$\pm$2.7 & -- \\
           & 1667.359  & --     & --           & --       & --       & --   & --             & -- \\
\hline\noalign{\smallskip}
\multicolumn{9}{l}{Comet C/2023 A3 (Tsuchinshan-ATLAS)} \\
\hline\noalign{\smallskip}
2024-10-13 & 1665.402  & 0.61$\pm$0.11   & 2.83$\pm$0.26  & -192.11$\pm$15.08  & 43.91  & 4.4 & -47.2$\pm$2.6 & -- \\
           & 1667.359  & 0.74$\pm$0.06   & 2.32$\pm$0.15  & -325.54$\pm$18.03  & 50.17  & 6.5 & -65.6$\pm$2.8 & 7.5$\pm$0.3 \\
           & 1720.530  & 0.47$\pm$0.15   & 1.93$\pm$0.36  & -100.98$\pm$16.18  & 37.79  & 2.7 & -17.0$\pm$1.9 & -- \\
\hline\noalign{\smallskip}
2024-10-14 & 1665.402  & 1.12$\pm$0.10   & 2.51$\pm$0.24  & -196.92$\pm$16.12  & 43.98  & 4.5 & -42.9$\pm$2.5 & -- \\
           & 1667.359  & 1.22$\pm$0.08   & 2.46$\pm$0.19  & -246.30$\pm$16.20  & 44.98  & 5.5 & -52.5$\pm$2.6 & 6.8$\pm$0.3 \\
\hline\noalign{\smallskip}
\end{tabular}\\
Notes: Column (1): Observational days with \ce{OH} detection; Column (2): \ce{OH} rest frequency;
Column (3)-(5): the Gaussian-fitted center velocity with respect to the comet nucleus, the FWHM line width,
and peak antenna temperature, respectively; Column (6)-(7): the rms noise of the baseline and peak intensity signal-to-noise ratio, respectively; Column (8): the integrated area; Column (9): the calculated \ce{OH} production rates for the 1667 MHz line.
\end{table*}

\section{Results and Analysis}\label{sec.res}

\subsection{The OH excitation}\label{subsec.exc}
The excitation of the OH 18-cm lines in cometary coma is governed by the interplay between solar ultraviolet (the Fraunhofer lines) pumping and collisional processes in the inner coma \citep{Despois1981, Schleicher1988}. The population inversion between the $\Lambda$-doublet sublevels, parameterized by the inversion factor $i$, determines whether the 18-cm lines appear in emission or absorption against the cosmic microwave background \citep{Schleicher1988}. Due to the complexity of the Fraunhofer lines, the pumping rate of the \ce{OH} lines strongly depends on the heliocentric radial velocity of the comet (also known as the Swings effect), which will affect the observed spectra \citep{Mies1974}.

For comet 12P, during our five epochs of observations, the detected OH lines at 1665 and 1667 MHz appeared in absorption, indicating a negative inversion factor during the observing epoch. On April 17, 2024, the line intensity ratio $I_{1667}/I_{1665} = 2.9\pm0.3$ was higher than the theoretical ratio of 1.8 under LTE condition, suggesting a further deviation from LTE. By May 10, 2024, however, this line intensity ratio had decreased to $1.6\pm0.2$, and only the 1665 MHz line was detected on the following third day. It was also noticed that the intensity of 1667 MHz line decreased between these two epochs, consistent with its non-detection in subsequent observations. Excluding the two low-SNR \ce{OH} detections obtained on May 10 and 13, the overall low detection rate was largely attributable to unfavorable heliocentric radial velocity during our observations. On April 17, the comet had a heliocentric radial velocity of $\dot{r} = -3.1$ km s$^{-1}$, corresponding to an negative inversion factor of -0.164. By April 24, the inversion factor had increased to +0.05, indicating a transition from absorption to emission in the \ce{OH} excitation. Over the following days, the inversion factor varied between -0.2 and -0.09, suggesting that the \ce{OH} excitation was approaching another transitional state. Similar transitions between emission and absorption in \ce{OH} excitation were also reported for comet C/2025 A6 by \citet{Jiang2026}.

This transitional behavior was also observed for comet A3 during our six epochs of observations. During the first two epochs, namely October 3 and 4, 2024, the corresponding inversion factors were -0.004 and +0.063, respectively, resulting in non-detections. Thereafter, the inversion factor evolved from -0.254 (October 13) and -0.228 (October 14) toward -0.158 (the last two epochs), as the inversion approached the next transition. Consequently, only the observations on October 13 and 14 yielded the remarkably high SNR results. 

The strength of the two \ce{OH} satellite lines is weak under the LTE conditions. Their detection under non-LTE conditions is also rare \citep{Bockelee-Morvan1992}. For the \ce{OH} satellite line at 1721 MHz, it was first detected in comet C/1978 H1 (Meier) at a 5$\sigma$ level, followed by a second detection in comet Levy (1990c) at an 8$\sigma$ level by \citet{Bockelee-Morvan1992}. \citet{Gerard1998} also reported the two 1612 and 1721 MHz lines for comet C/1996 B2 (Hyakutake) after averaged the spectra over a range of days. In recent years, \citet{Smith2021} reported a 1$\sigma$ detection of the 1721 MHz line for comet C/2020 F3 (NEOWISE) using the Arecibo radio telescope. In our observations of comet A3, the SNR for the 1721 MHz line is 2.7$\sigma$. To further verify its identification, we compared consistency of the line profiles among the 1665, 1667, and 1721 MHz lines, and checked the polarization difference of the spectra. We noticed that the line center of the 1721 MHz line is slightly offset by 0.27 km s$^{-1}$ with respect to the 1667 MHz line. This discrepancy was also found by \citep{Smith2021}. For the polarizations, it can be seen in Fig.~\ref{OH_A3_1721_pola} that both the 1665 and 1667 MHz lines showed absorption in the XX and YY polarizations, whereas for the 1721 MHz line, absorption was only seen in the YY polarization spectra. It was noted that the blueshifted broad line wing in the XX polarization of the 1667 MHz line is due to noise fluctuations of the spectra at the source OFF position. For the other lines, no radio frequency interference (RFI) was present in the velocity range of the signal during the observations. We noted that the local RFI is usually polarized if presented, and that the noise diode temperatures vary differently for the two polarizations \citep{Zhang2023}. One of the reasons for the absence of absorption in the XX polarization for the 1721 MHz line may be cancellation caused by RFI emission. Therefore, based on the above reasons, we denote the 1721 MHz absorption line as a tentative identification. The line intensity ratios of 1721:1667:1665 were 0.069:1:0.582 for comet Levy (1990c). As a comparison, the ratios were 0.26:1:0.72 for comet A3 in this study.

For comets, the polarization can be caused by the scattering of solar radiation by cometary grains in the coma or jets \citep{Levasseur-Regourd1999}, which has been reported in the optical (\citep{Eaton1988, Maslov2021}) and near-infrared (\citep{Lim2025}) observations. At the radio wavelengths, \citet{Cordes1990} found no strong evidence of polarization effect for the 1667 MHz line, and provided only a weak upper limit on the linear polarization of 10 percent for comet P/Halley with Arecibo telescope. Further verification of the polarizations and satellite lines for other bright comets could provide valuable insights.

\subsection{The OH production rates}
We calculated the \ce{OH} production rate using the equation from \citet{Bockelee-Morvan1990}, which can be expressed as:
\begin{equation}
Q_\mathrm{OH} = 2.33\times10^{34}\frac{\Delta^2 S}{\tau_\mathrm{OH} T_\mathrm{bg}}\frac{1}{i}\frac{1}{f},
\end{equation}
where $S$ is the integrated area of the \ce{OH} line in $\mathrm{Jy~km~s^{-1}}$, and $\Delta$ is the Earth-comet distance in AU. $T_\mathrm{bg}$ and $i$ are the background temperature in K and inversion factor, respectively. $f$ and $\tau_\mathrm{OH}$ are the fraction of \ce{OH} at fluorescence equilibrium and the \ce{OH} lifetime, respectively. $\tau_\mathrm{OH} = \tau_\mathrm{OH(1\,AU)} r_h^2$, in which $r_\mathrm{h}$ is the comet's heliocentric distance in AU. $f$ depends on the beam size and quenching radius of the gas coma, as discussed below. The inversion factor $i$ depends on the heliocentric radial velocity of the comet, which is taken from \citet{Schleicher1988}. We further assumed $T_\mathrm{bg}=3$ K, as the two comets were far away from the Galactic plane during the observations, and $\tau_\mathrm{OH(1\,AU)} = 1.1\times10^{5}$ s \citep{Crovisier2002}.

For the calculation of the \ce{OH} production rate, we found that the small beam size and the quenching effect of the gas coma contribute significantly to the estimation of the $f$ factor, as explained below.

First, small beam size could lead to the loss of a fraction of the \ce{OH} flux density. The beam size of FAST is 2.47$^\prime$ at 1667 MHz, corresponding to radii of 85,989 and 26,872 km from the center of the comet nucleus for 12P and A3 during the observations, respectively. For comparison, assuming an expansion velocity of 0.9 km s$^{-1}$, the scale-lengths of \ce{OH} are 63,360 and 35,640 km for 12P and A3 during the observations, respectively. We can see that the scale-length is comparable to the beam radius for comet 12P, and is even larger than the beam radius for comet A3. This phenomenon was reported for comet P/Halley observed with the Arecibo radio telescope \citep{Cordes1990}, which concluded that only roughly 3--10 percent of the \ce{OH} flux was sampled.

Second, the neglect of the collisional quenching effect could underestimate the \ce{OH} production rate. A factor $f=1$ of \ce{OH} at fluorescence equilibrium corresponds to no quenching effect, which is usually the case for an unresolved \ce{OH} coma \citep{Bockelee-Morvan1990}. \citet{Smith2021} reported an \ce{OH} production rate of 3.6$\times$10$^{28}$ molecules s$^{-1}$ for comet C/2020 F3 (NEOWISE) with the Arecibo telescope, which was lower than the values reported for the same comet observed with the Nan\c{c}ay Radio Telescope \citep{Drozdovskaya2023}. \citet{Drozdovskaya2023} further showed that, by considering the effect of collisional quenching, the derived updated \ce{OH} production rate increased by a factor of 4--8.

In this study, we estimated the $f$ factor following the procedure presented in \citet{Li2025, Jiang2026}. After considering the weighting of the beam size and the collisional quenching effect, $f$ can be expressed as
\begin{equation}
f = \frac{\int_0^{+\infty} N^{\prime}(\rho) \omega(\rho) \rho~\mathrm{d}\rho}{\int_0^{+\infty} N(\rho) \rho~\mathrm{d}\rho},
\end{equation}
where $\rho$ is the projected nucleocentric distance, $\omega(\rho)$ is the beam-weighting Gaussian function, and $N(\rho)$ and $N^{\prime}(\rho)$ are the total \ce{OH} column density distribution and the column density distribution after subtracting the collisional quenching sphere, respectively.

The \ce{OH} column density can be obtained by integrating the \ce{OH} number density distribution \citep{Despois1981}, which can be expressed as
\begin{equation}
n_{\mathrm{OH}}(r) = \frac{Q_\mathrm{p}}{4\pi r^2 v_{\mathrm{OH}}} \frac{L_{\mathrm{OH}}}{L_{\mathrm{OH}} - L_{\mathrm{p}}} (e^{-r/L_{\mathrm{OH}}} - e^{-r/L_{\mathrm{p}}}),
\end{equation}
where $r$ is the nucleocentric distance, and $Q_\mathrm{p}$ is the parent production rate, which we assumed to be that of \ce{H2O} with $Q_\mathrm{H_2O} = 1.1Q_\mathrm{OH}$. $v_{\mathrm{OH}}$ is the ejection velocity of \ce{OH}, for which a value of $v_{\mathrm{OH}} = 0.95$ km s$^{-1}$ was assumed. $L_{\mathrm{OH}}$ and $L_{\mathrm{p}}$ are the scale-lengths of the daughter and parent species. For \ce{H2O}, $L_{\mathrm{p}} = v_{\mathrm{p}} \tau_\mathrm{p(1\,AU)} r_\mathrm{h}^2$, where $\tau_\mathrm{p(1\,AU)} = 8.5\times10^4$ s at 1 AU \citep{Crovisier2002} and $v_{\mathrm{p}}$ was obtained from the fitted profile.

As for the collisional quenching sphere, we followed the expression presented in \citet{Gerard1998, Li2025}
\begin{equation}
r_{\mathrm{q}} = r_{\mathrm{q}}^* r_\mathrm{h} \sqrt{Q_{\mathrm{OH}}/10^{29}},
\end{equation}
where $r_{\mathrm{q}}$ is the quenching radius in km, and $r_{\mathrm{q}}^* = 47,000$ km is the characteristic quenching radius for an \ce{OH} production rate of $10^{29}$ molecules s$^{-1}$ at a heliocentric distance of 1 AU. This characteristic value is derived from one or more template comet observations. Within $r_{\mathrm{q}}^*$, collisions suppress the \ce{OH} maser excitation, and we therefore assume the inversion factor to be zero.

Finally, an iterative procedure was used to estimate the $f$ factor and $Q_{\mathrm{OH}}$. However, for comet 12P on April 17, the iterative procedure did not converge with the default value of $r_{\mathrm{q}}^* = 47,000$ km. During the iteration, the derived quenching radius progressively increased until it exceeded the \ce{OH} scale length, preventing the solution from converging. The non-convergence may be result from the high \ce{OH} production rate on that date, which in turn may be related to the outburst event (see Section \ref{disc.outburst} for the discussion). We therefore searched for the maximum $r_{\mathrm{q}}^*$ that allowed convergence, using a grid with a step of 1,000 km. The reduced characteristic quenching radius found was 44,000 km. At convergence, for comet 12P, the derived quenching radii were 69,634 and 42,217 km on April 17 and May 10, respectively. For comet A3, these values were 23,250 and 22,938 km on October 13 and 14, respectively. The derived $f$ factor was 0.3 for comet 12P and 0.2 for comet A3. 
For three of the four measurements, the derived quenching radii were 0.81--0.92 times the half-beam size. The exception is comet 12P on May 10, for which $r_q$ was only 0.50 times the half-beam size, possibly due to the lower \ce{OH} production rate or the greater uncertainty associated with the low-SNR spectrum on that date. The calculated \ce{OH} production rates for the 1667 MHz line are presented in the last column of Table \ref{line_profile}. Since $r_{\mathrm{q}}^*$ is empirically determined and has been estimated to range from 37,000 to 50,000 km \citep{Gerard1998}, we tested the sensitivity of the derived $Q_{\mathrm{OH}}$ to this parameter. Adopting $r_{\mathrm{q}}^* = 44,000$ km changes $Q_{\mathrm{OH}}$ by less than 9 percent, while $r_{\mathrm{q}}^* = 37,000$ km produces a maximum difference of 27 percent, occurring for comet 12P on April 17.

\section{Discussion}\label{sec.disc}

\subsection{Comet 12P and comparison with other Halley-type comets}\label{disc.outburst}
In this section we focus on the \ce{OH} variations at different heliocentric distances for comet 12P, and give a comparison between comet 12P and three other well-studied Halley-type comets (1P/Halley, 13P/Olbers, and 109P/Swift-Tuttle) based on their \ce{OH} production rates in order to find differences among them.

\begin{figure*}
\centering
\includegraphics[scale=0.4]{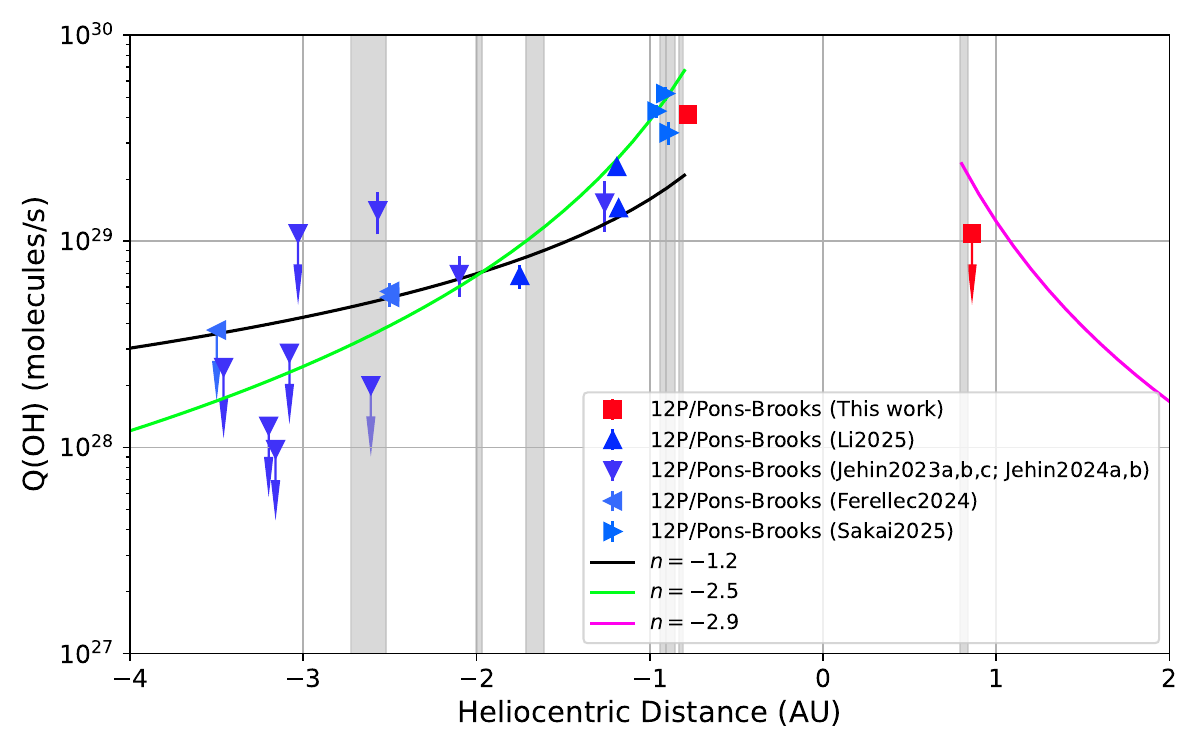}
\caption{The \ce{OH} production rates as a function of heliocentric distance for comet 12P, along with the values for this comet reported in the literature. The negative x-axis values represent the pre-perihelion heliocentric distance. The black line is the slope of the power-law fit from \citet{Li2025}, the green and pink lines are the slopes of the power-law fit from \citet{Combi2025} for pre-perihelion and post-perihelion, respectively. The gray vertical bands are the outburst durations indicated by the Lyman-$\alpha$ observations \citep{Combi2025}.}
\label{Q_OH_12P}
\end{figure*}

Figure \ref{Q_OH_12P} shows the calculated \ce{OH} production rates for comet 12P as a function of heliocentric distance in this study, which also includes results from optical observations \citep{Jehin2023a, Jehin2023b, Jehin2023c, Jehin2024a, Jehin2024b, Ferellec2024} and from radio observations \citep{Li2025, Sakai2025} for this comet, which were conducted before the comet's perihelion.

The \ce{OH} production rates for comet 12P before perihelion can reach as high as $5\times10^{29}$ molecules s$^{-1}$ \citep{Sakai2025}. These high production rates coincided with outburst events indicated by the SOHO/SWAN Lyman-$\alpha$ observations \citep{Combi2025}. In contrast, no outburst was reported during this study. \citet{Li2025} reported a power-law fit of the \ce{OH} production rates as a function of heliocentric distance with a slope of -1.2. The hydrogen Lyman-$\alpha$ observations from \citet{Combi2025} showed that the power-law slopes derived by the water production rates are -2.5 and -2.9 for pre-perihelion and post-perihelion, respectively. These power-law fits are also overlaid on Fig.~\ref{Q_OH_12P}, assuming a yield of 0.82 from water to \ce{OH} \citep{Combi2025}.

One of the most remarkable aspects of comet 12P during its 2024 apparition was the series of outbursts. Two optical outbursts occurred after the observation of \citet{Li2025}, who found a 37 percent decrease in \ce{OH} production between observations on March 2 and 3 that may be related to the outburst. \citet{Sakai2025} showed a comparison of the \ce{OH} productions with the optical outbursts, in which no outbursts were correlated with their observations, but they also found a 36 percent decrease in \ce{OH} production between March 27 and 29. To examine the effect of outbursts on \ce{OH} production, we also overlaid the durations of 7 outburst events indicated by the SOHO/SWAN Lyman-$\alpha$ observations \citep{Combi2025} on Fig.~\ref{Q_OH_12P}, shown by the gray vertical bands. For the outbursts indicated by the Lyman-$\alpha$ observations, the three observation days from \citet{Sakai2025} correspond to the pre-outburst phase, the start of the outburst (March 27), and the peak of the outburst (March 29), and the maximum \ce{OH} production coincided with the start of the outburst. No outbursts indicated by the Lyman-$\alpha$ observations were recorded during observations of \citet{Li2025}. It was also noticed that both optical and Lyman-$\alpha$ observations recorded an outburst on November 14, 2023, and there was a 7-fold increase in \ce{OH} production between November 12 and 15, 2023, as shown by optical observations \citet{Jehin2023c}. For our observation on April 17, it was 9 days after the start of the last outburst (April 8), while our observation on May 10 was 10 days after the last outburst (April 30). Therefore, no meaningful conclusions could be drawn from our observations. The correlation between \ce{OH} production and the outburst indicated by the Lyman-$\alpha$ observations is reasonable, as both \ce{H} and \ce{OH} are the products of the photodissociation of \ce{H2O}, whereas the correlation between \ce{OH} production and optical outburst needs further verification.

\begin{figure*}
\centering
\includegraphics[scale=0.4]{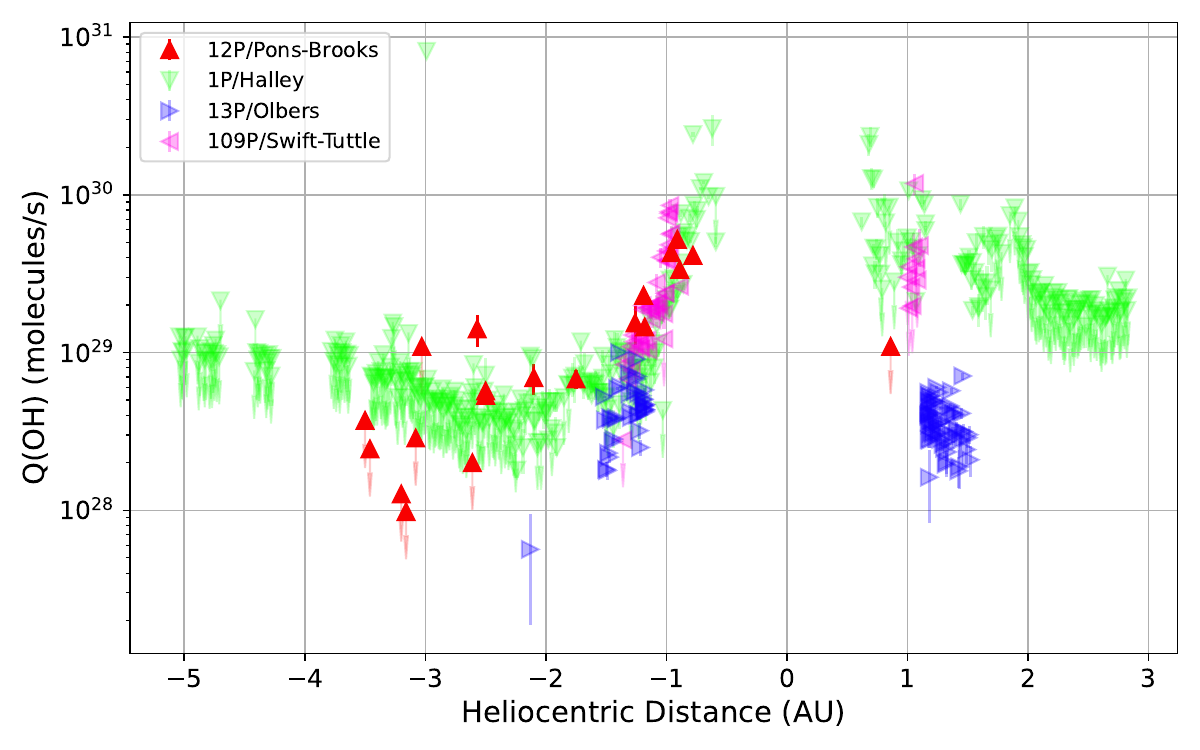}
\caption{The \ce{OH} production rates as a function of heliocentric distance for comet 12P, along with the values for the three other Halley-type comets reported in the literature. The negative x-axis values represent the pre-perihelion heliocentric distance.}
\label{Q_OH_12P_HTCs}
\end{figure*}

To evaluate the differences between comet 12P and other Halley-type comets, Fig.~\ref{Q_OH_12P_HTCs} shows the variations of \ce{OH} production as a function of heliocentric distance for comet 12P and three well-studied Halley-type comets, 1P/Halley, 13P/Olbers, and 109P/Swift-Tuttle. Comet 1P/Halley was monitored by \citet{Crovisier2002} using the Nan\c{c}ay Radio Telescope over a long time range; thus the \ce{OH} productions span a large range of heliocentric distances. However, many of those derived \ce{OH} productions are upper-limit values. For comet 13P/Olbers, most of the \ce{OH} productions were taken from SOHO/SWAN Lyman-$\alpha$ observations \citep{Combi2025} using a yield of 0.82 for \ce{H2O} to \ce{OH}, and two values were taken from optical observations by \citet{Jehin2024b, Jehin2024c}. These two values are smaller by a factor of a few, but should have no impact on our general conclusions. For comet 109P/Swift-Tuttle, the \ce{OH} productions were also taken from \citet{Crovisier2002}.

From Fig.~\ref{Q_OH_12P_HTCs} we can see that before perihelion, these four Halley-type comets have similar yields in their \ce{OH} productions. However, after perihelion, some divergences appear: the \ce{OH} productions are systematically lower by a factor of several for comet 13P/Olbers than those for 1P/Halley and 109P/Swift-Tuttle. Individually, the \ce{OH} productions are similar between pre- and post-perihelion for comets 13P/Olbers and 109P/Swift-Tuttle, whereas for comet 1P/Halley, the \ce{OH} productions are higher post-perihelion than pre-perihelion. For comet 12P, there is only one value in the post-perihelion, and it appears to fill the gap in the distribution of \ce{OH} productions for these Halley-type comets. Moreover, for these four Halley-type comets, although they differ in average production rates, they have very similar behavior as they move through perihelion within the inner 1--1.5 AU region. Therefore, generally speaking, comet 12P is a typical Halley-type comet.

\begin{figure*}
\centering
\includegraphics[scale=0.4]{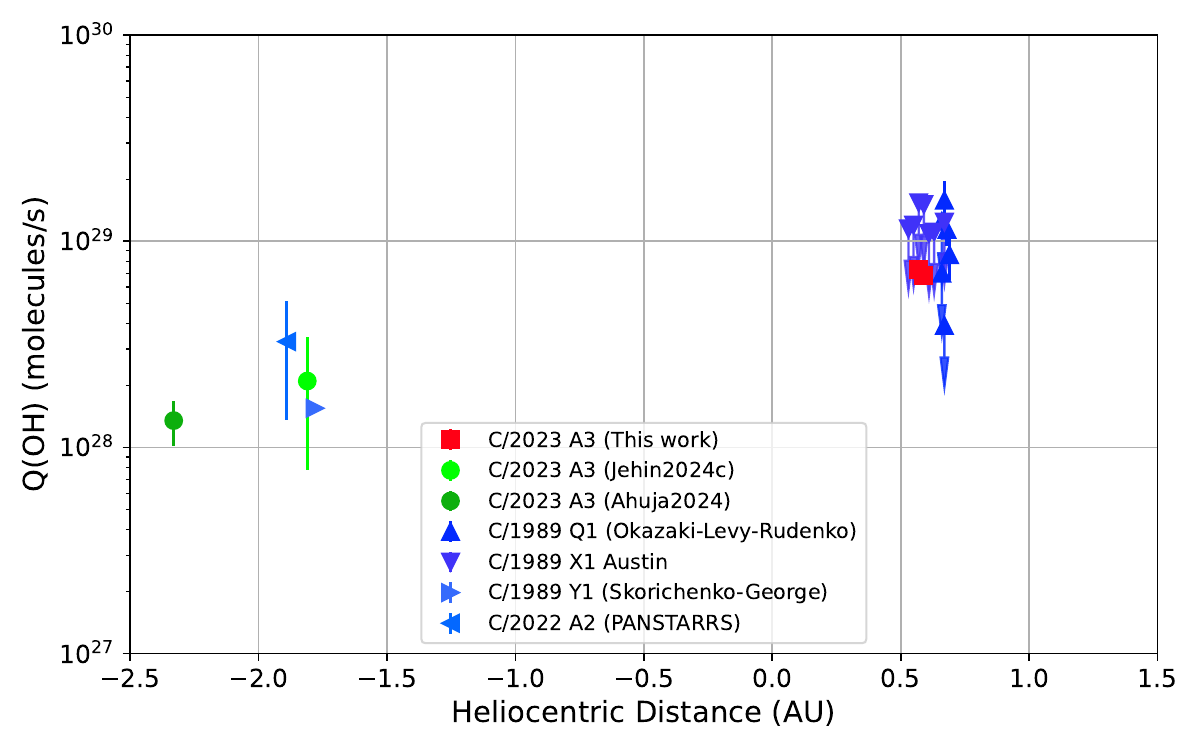}
\caption{The \ce{OH} production rates as a function of heliocentric distance for comet A3, along with the values for the four other dynamically new comets reported in the literature. The negative x-axis values represent the pre-perihelion heliocentric distance.}
\label{Q_OH_A3}
\end{figure*}

\subsection{Comet A3 and comparison with other dynamically new Oort Cloud comets}
Unlike comet 12P, comet A3 had a much greater visual magnitude during its approach to the Sun, making it one of the brightest comets observed in recent decades. In contrast, fewer species were reported for comet A3 before its perihelion compared with comet 12P. While the exceptional brightness was partly enhanced by favorable forward-scattering geometry near perihelion \citep{Moreno2025}, it has also been suggested that this comet possessed a relatively high dust-to-gas ratio and was classified as a carbon-depleted comet \citep{Ahuja2024, Tang2024, Cambianica2025}.

For comet A3, \citet{Ahuja2024} calculated a water production rate of $(1.50\pm0.37) \times 10^{28}$ molecules s$^{-1}$ at a pre-perihelion distance of 2.33 AU from optical observations. \citet{Jehin2024c} reported an optical \ce{OH} production rate of $(2.1\pm1.3) \times 10^{28}$ molecules s$^{-1}$ when the comet was at a pre-perihelion heliocentric distance of 1.81 AU. Other values were not yet reported in the literature for this comet. In Fig.~\ref{Q_OH_A3}, we compared the calculated \ce{OH} production rates for this comet with those of four other dynamically new comets obtained from the literature at similar heliocentric distances. Specifically, these are comet C/1989 Q1 (Okazaki-Levy-Rudenko) \citep{Crovisier2002}, C/1989 X1 Austin \citep{Crovisier2002}, C/1989 Y1 (Skorichenko-George) \citep{AHearn1995}, and C/2022 A2 (PANSTARRS) \citep{Jehin2022}. We can see that the \ce{OH} production for comet A3 is roughly consistent with that for the other comets.

However, there are also differences among them. Comet C/1989 Q1 (Okazaki-Levy-Rudenko) is gas-rich with a typical gas component \citep{Winiarski1992, Kolokolova2007} according to the taxonomic classes based on the $\log[Q(\rm C_2)/Q(\rm CN)]$ ratio from \citet{AHearn1995}. Comet C/1989 X1 Austin prefers a typical gas component but shows significant variation in the dust-to-gas ratio \citep{Joshi1992, Kolokolova2007}. Comets C/1989 Y1 (Skorichenko-George) and C/2022 A2 (PANSTARRS) both have typical gas components and tend toward dust-rich properties based on the $\log(A(\theta)f\rho / Q({\rm OH}))$ ratio \citep{AHearn1995, Jehin2022, Jehin2023a}.

Due to the lack of available data for comet A3 from the same epochs as our observations, no constraints on its properties could be drawn from this study. Based on the high dust-to-gas ratio and carbon-depleted nature of comet A3, we briefly discuss its implications. Although the dust-to-gas ratio itself may not correlate with a comet's dynamical age, there is increasing evidence that compositional classes of comets are related to their formation environments \citep{AHearn1995, Mumma2011}. A relatively high dust-to-gas ratio may be consistent with formation in a region where dust coagulation into pebbles was particularly efficient, volatile depletion occurred before planetesimal formation, or refractory material was preferentially incorporated into planetesimals. However, the observed high dust-to-gas ratio does not necessarily reflect primordial composition. Evolutionary effects, such as preferential depletion of near-surface volatiles or unusually efficient dust production, could produce similar observational signatures. Distinguishing between these possibilities would require combining the OH measurements with independent constraints on dust production, volatile abundances, and dust properties, which is beyond the scope of the present paper.

\section{Summary}\label{sec.sum}
In this paper, we have presented radio observations of the OH 18-cm lines in comets 12P/Pons-Brooks and C/2023 A3 (Tsuchinshan-ATLAS) using the Five-hundred-meter Aperture Spherical radio Telescope (FAST). Due to unfavorable heliocentric radial velocity during the observations, low \ce{OH} detection rates were achieved. Despite this, we securely detected the OH absorption lines at 1665 and 1667 MHz in both comets. Although the small beam size of FAST results in the loss of a fraction of flux, there is an advantage in that the small beam size could resolve the gas coma. Combined with the FAST's high sensitivity, weak \ce{OH} excitation can be detected. Therefore, studies of the collisional quenching region and local/transient activity in the inner gas coma could be an opportunity for FAST. On the other hand, 18-cm \ce{OH} radio observations provide an important complement to optical and infrared observations by offering an independent probe of water production over a wide range of excitation conditions. Combining radio \ce{OH} measurements with modern multi-wavelength observations can provide much stronger constraints on cometary water production and activity.

The main conclusions of this study are summarized as follows:

\begin{enumerate}
\item For comet 12P/Pons-Brooks, observations on April 17, 2024 (pre-perihelion) yielded an \ce{OH} production rate of $(4.12\pm0.19) \times 10^{29}$ molecules s$^{-1}$. Upper limited detections on May 10 and 13 suggested declining $\ce{OH}$ production rates after the post-perihelion.

\item For comet C/2023 A3 (Tsuchinshan-ATLAS), we obtained six epochs of data after the post-perihelion, with high SNR detections only on October 13 and 14, 2024, yielding an \ce{OH} production rate of $(7.5\pm0.3) \times 10^{28}$ molecules s$^{-1}$ and $(6.8\pm0.3) \times 10^{28}$ molecules s$^{-1}$, respectively.

\item Tentative identification of the 1721 MHz satellite line in comet C/2023 A3 (Tsuchinshan-ATLAS).

\item For comet 12P/Pons-Brooks, the \ce{OH} production rates follow the power-law distribution well, consistent with other observations of this comet at different heliocentric distances and epochs. By comparing the variation of \ce{OH} production for this comet with that for three other Halley-type comets, we find that comet 12P/Pons-Brooks is a typical Halley-type comet.

\item For comet C/2023 A3 (Tsuchinshan-ATLAS), although the derived \ce{OH} production rates appear relatively modest compared with its exceptional brightness, they are consistent with those from other dynamically new Oort Cloud comets at comparable heliocentric distances.
\end{enumerate}

\begin{acknowledgments}
We thank the reviewer for the helpful suggestions that have improved and clarified the manuscript.
The authors thank Qingliang Yang, Chun Sun, and Peng Jiang at FAST Operation Center for assisting in the development of the special observation mode for moving target observations.
L.-F.C.~thanks Dongyue Jiang and Chuan-Peng Zhang for the useful discussions on the data processing and \ce{OH} excitation.
C.-W.T.~acknowledges the support from National Natural Science Foundation of China (NSFC) No.~12588202. 
Z.W.~acknowledges the support from Natural Science Foundation of Xinjiang Uygur Autonomous Region No.~2025D01E62.
J.-Y.L. acknowledges the support by the 2024 Xinjiang Autonomous Region Tianchi Talent Program.

\end{acknowledgments}

%
\facilities{FAST}

\software{astropy \citep{astropy2013, astropy2018, astropy2022}, numpy \citep{numpy2020}, scipy \citep{scipy2020}, matplotlib \citep{mpl2007}, pandas \citep{pandas2010,pandas2025}, specutils \citep{specutils2019}
}



\begin{thebibliography}{90}
\bibitem[\protect\citeauthoryear{A'Hearn et al.}{1995}]{AHearn1995} A'Hearn M.~F., Millis R.~C., Schleicher D.~O., Osip D.~J., Birch P.~V., 1995, Icar, 118, 223. doi:10.1006/icar.1995.1190
\bibitem[\protect\citeauthoryear{Ahuja et al.}{2024}]{Ahuja2024} Ahuja G., Aravind K., Sahu D., Jehin E., Donckt M.~V., Hmiddouch S., Ganesh S., et al., 2024, ATel, 16637, 1
\bibitem[\protect\citeauthoryear{Altwegg}{2022}]{Altwegg2022} Altwegg K., 2022, PhT, 75, 34. doi:10.1063/PT.3.4920
\bibitem[\protect\citeauthoryear{Astropy Collaboration et al.}{2013}]{astropy2013} Astropy Collaboration, Robitaille T.~P., Tollerud E.~J., Greenfield P., Droettboom M., Bray E., Aldcroft T., et al., 2013, A\&A, 558, A33. doi:10.1051/0004-6361/201322068
\bibitem[\protect\citeauthoryear{Astropy Collaboration et al.}{2018}]{astropy2018} Astropy Collaboration, Price-Whelan A.~M., Sip{\H{o}}cz B.~M., G{\"u}nther H.~M., Lim P.~L., Crawford S.~M., Conseil S., et al., 2018, AJ, 156, 123. doi:10.3847/1538-3881/aabc4f
\bibitem[\protect\citeauthoryear{Astropy Collaboration et al.}{2022}]{astropy2022} Astropy Collaboration, Price-Whelan A.~M., Lim P.~L., Earl N., Starkman N., Bradley L., Shupe D.~L., et al., 2022, ApJ, 935, 167. doi:10.3847/1538-4357/ac7c74
\bibitem[\protect\citeauthoryear{Astropy-Specutils Development Team}{2019}]{specutils2019} Astropy-Specutils Development Team, 2019, ascl.soft. ascl:1902.012
\bibitem[\protect\citeauthoryear{Biraud et al.}{1974}]{Biraud1974} Biraud F., Bourgois G., Crovisier J., Fillit R., Gerard E., Kazes I., 1974, A\&A, 34, 163
\bibitem[\protect\citeauthoryear{Biver et al.}{2024a}]{Biver2024a} Biver N., Bockel{\'e}e-Morvan D., Handzlik B., Sandqvist A., Boissier J., Drozdovskaya M.~N., Moreno R., et al., 2024, A\&A, 690, A271. doi:10.1051/0004-6361/202450921
\bibitem[\protect\citeauthoryear{Biver et al.}{2024b}]{Biver2024b} Biver N., Boissier J., Bockel{\'e}e-Morvan D., Crovisier J., Moreno R., Cordiner M., Lis D.~C., et al., 2024, EPSC, EPSC2024-371. doi:10.5194/epsc2024-371
\bibitem[\protect\citeauthoryear{Bockel{\'e}e-Morvan, Crovisier, \& G{\'e}rard}{1990}]{Bockelee-Morvan1990} Bockel{\'e}e-Morvan D., Crovisier J., G{\'e}rard E., 1990, A\&A, 238, 382
\bibitem[\protect\citeauthoryear{Bockelee-Morvan et al.}{1992}]{Bockelee-Morvan1992} Bockelee-Morvan D., Colom P., Crovisier J., Gerard E., Bourgois G., 1992, acm..proc, 73
\bibitem[\protect\citeauthoryear{Bockel{\'e}e-Morvan et al.}{2015}]{Bockelee-Morvan2015} Bockel{\'e}e-Morvan D., Calmonte U., Charnley S., Duprat J., Engrand C., Gicquel A., H{\"a}ssig M., et al., 2015, SSRv, 197, 47. doi:10.1007/s11214-015-0156-9
\bibitem[\protect\citeauthoryear{Cambianica et al.}{2025}]{Cambianica2025} Cambianica P., Munaretto G., Cremonese G., Mura A., La Forgia F., Bizzocchi L., Lazzarin M., et al., 2025, P\&SS, 261, 106102. doi:10.1016/j.pss.2025.106102
\bibitem[\protect\citeauthoryear{Chambers}{2023}]{Chambers2023} Chambers J., 2023, ApJ, 944, 127. doi:10.3847/1538-4357/aca96f
\bibitem[\protect\citeauthoryear{Chen et al.}{2024}]{Chen2024} Chen L.-F., Tsai C.-W., Li J.-Y., Yang B., Li D., Duan Y., Hsia C.-H., et al., 2024, RAA, 24, 105008. doi:10.1088/1674-4527/ad7823
\bibitem[\protect\citeauthoryear{Combi et al.}{2025}]{Combi2025} Combi M.~R., M{\"a}kinen T., Bertaux J.-L., Qu{\'e}merais E., Ferron S., 2025, PSJ, 6, 306. doi:10.3847/PSJ/ae2475
\bibitem[\protect\citeauthoryear{Cordes et al.}{1990}]{Cordes1990} Cordes J.~M., Falchi A., Lewis B.~M., Tofani G., Terzian Y., 1990, AJ, 100, 1655. doi:10.1086/115625
\bibitem[\protect\citeauthoryear{Crovisier et al.}{2002}]{Crovisier2002} Crovisier J., Colom P., G{\'e}rard E., Bockel{\'e}e-Morvan D., Bourgois G., 2002, A\&A, 393, 1053. doi:10.1051/0004-6361:20020673
\bibitem[\protect\citeauthoryear{Despois et al.}{1981}]{Despois1981} Despois D., Gerard E., Crovisier J., Kazes I., 1981, A\&A, 99, 320
\bibitem[\protect\citeauthoryear{Drozdovskaya et al.}{2023}]{Drozdovskaya2023} Drozdovskaya M.~N., Bockel{\'e}e-Morvan D., Crovisier J., McGuire B.~A., Biver N., Charnley S.~B., Cordiner M.~A., et al., 2023, A\&A, 677, A157. doi:10.1051/0004-6361/202346402
\bibitem[\protect\citeauthoryear{Eaton, Scarrott, \& Warren-Smith}{1988}]{Eaton1988} Eaton N., Scarrott S.~M., Warren-Smith R.~F., 1988, Icar, 76, 270. doi:10.1016/0019-1035(88)90072-3
\bibitem[\protect\citeauthoryear{Ehrenfreund \& Charnley}{2000}]{Ehrenfreund2000} Ehrenfreund P., Charnley S.~B., 2000, ARA\&A, 38, 427. doi:10.1146/annurev.astro.38.1.427
\bibitem[\protect\citeauthoryear{Ferellec et al.}{2024}]{Ferellec2024} Ferellec L., Opitom C., Donaldson A., Fynbo J.~P.~U., Kokotanekova R., Kelley M.~S.~P., Lister T., 2024, MNRAS, 534, 1816. doi:10.1093/mnras/stae2189
\bibitem[\protect\citeauthoryear{G{\'e}rard et al.}{1998}]{Gerard1998} G{\'e}rard E., Crovisier J., Colom P., Biver N., Bockel{\'e}e-Morvan D., Rauer H., 1998, P\&SS, 46, 569. doi:10.1016/S0032-0633(97)00197-9
\bibitem[\protect\citeauthoryear{Gritsevich, Weso{\l}owski, \& Castro-Tirado}{2025}]{Gritsevich2025} Gritsevich M., Weso{\l}owski M., Castro-Tirado A.~J., 2025, MNRAS, 538, 470. doi:10.1093/mnras/staf219
\bibitem[\protect\citeauthoryear{Harris et al.}{2020}]{numpy2020} Harris C.~R., Millman K.~J., van der Walt S.~J., Gommers R., Virtanen P., Cournapeau D., Wieser E., et al., 2020, Natur, 585, 357. doi:10.1038/s41586-020-2649-2
\bibitem[\protect\citeauthoryear{Huebner, Keady, \& Lyon}{1992}]{Huebner1992} Huebner W.~F., Keady J.~J., Lyon S.~P., 1992, Ap\&SS, 195, 1. doi:10.1007/BF00644558
\bibitem[\protect\citeauthoryear{Hunter}{2007}]{mpl2007} Hunter J.~D., 2007, CSE, 9, 90. doi:10.1109/MCSE.2007.55
\bibitem[\protect\citeauthoryear{Jehin et al.}{2022}]{Jehin2022} Jehin E., Vander Donckt M., Manfroid J., Hmiddouch S., Moulane Y., Jabiri A., Benkhaldoun Z., 2022, ATel, 15822, 1
\bibitem[\protect\citeauthoryear{Jehin et al.}{2023a}]{Jehin2023a} Jehin E., Donckt M.~V., Hmiddouch S., Manfroid J., Jabiri A., Benkhaldoun Z., 2023, ATel, 16223, 1
\bibitem[\protect\citeauthoryear{Jehin et al.}{2023b}]{Jehin2023b} Jehin E., Vander Donckt M., Hmiddouch S., Manfroid J., Jabiri A., Benkhaldoun Z., 2023, ATel, 16282, 1
\bibitem[\protect\citeauthoryear{Jehin et al.}{2023c}]{Jehin2023c} Jehin E., Donckt M.~V., Hmiddouch S., Manfroid J., 2023, ATel, 16338, 1
\bibitem[\protect\citeauthoryear{Jehin et al.}{2024a}]{Jehin2024a} Jehin E., Donckt M.~V., Hmiddouch S., Manfroid J., 2024, ATel, 16408, 1
\bibitem[\protect\citeauthoryear{Jehin et al.}{2024b}]{Jehin2024b} Jehin E., Donckt M.~V., Hmiddouch S., Manfroid J., 2024, ATel, 16498, 1
\bibitem[\protect\citeauthoryear{Jehin et al.}{2024c}]{Jehin2024c} Jehin E., Donckt M.~V., Hmiddouch S., Manfroid J., 2024, ATel, 16705, 1
\bibitem[\protect\citeauthoryear{Jewitt \& Luu}{2025}]{Jewitt2025} Jewitt D., Luu J., 2025, AJ, 169, 338. doi:10.3847/1538-3881/add2ff
\bibitem[\protect\citeauthoryear{Jiang et al.}{2026}]{Jiang2026} Jiang D., Qian L., Guo M., Hao Q., Huang M., Jiang P., Liu H., et al., 2026, RAA, 26, 061002. doi:10.1088/1674-4527/ae561a
\bibitem[\protect\citeauthoryear{Joshi et al.}{1992}]{Joshi1992} Joshi U.~C., Sen A.~K., Deshpande M.~R., Chauhan J.~S., 1992, JApA, 13, 267. doi:10.1007/BF02702294
\bibitem[\protect\citeauthoryear{Knight et al.}{2024}]{Knight2024} Knight M.~M., Skiff B.~A., Schleicher D.~G., Spiro L.~G., Fernald I.~C., Guan B.~Y., Lininger L.~J., et al., 2024, ATel, 16508, 1
\bibitem[\protect\citeauthoryear{Kobayashi et al.}{2025}]{Kobayashi2025} Kobayashi H., Kawakita H., DiSanti M.~A., Bonev B.~P., Dello Russo N., Vervack R.~J., 2025, epsc.conf, 2025, EPSC-DPS2025-694. doi:10.5194/epsc-dps2025-694
\bibitem[\protect\citeauthoryear{Kolokolova et al.}{2007}]{Kolokolova2007} Kolokolova L., Kimura H., Kiselev N., Rosenbush V., 2007, A\&A, 463, 1189. doi:10.1051/0004-6361:20065069
\bibitem[\protect\citeauthoryear{Levasseur-Regourd}{1999}]{Levasseur-Regourd1999} Levasseur-Regourd A.-C., 1999, SSRv, 90, 163. doi:10.1023/A:1005250131509
\bibitem[\protect\citeauthoryear{Li et al.}{2025}]{Li2025} Li J., Shi X., Shi J., Ma Y., Yang B., Li J.-Y., Xing Z., et al., 2025, A\&A, 701, A204. doi:10.1051/0004-6361/202554867
\bibitem[\protect\citeauthoryear{Lim et al.}{2025}]{Lim2025} Lim B., Ishiguro M., Takahashi J., Akitakya H., Geem J., Bach Y.~P., Jin S., et al., 2025, ApJL, 983, L19. doi:10.3847/2041-8213/adc2f9
\bibitem[\protect\citeauthoryear{Liu, Hui, \& Liu}{2025}]{Liu2025} Liu B., Hui M.-T., Liu X., 2025, A\&A, 698, A95. doi:10.1051/0004-6361/202554632
\bibitem[\protect\citeauthoryear{Mandt et al.}{2024}]{Mandt2024} Mandt K.~E., Lustig-Yaeger J., Luspay-Kuti A., Wurz P., Bodewits D., Fuselier S.~A., Mousis O., et al., 2024, SciA, 10, eadp2191. doi:10.1126/sciadv.adp2191
\bibitem[\protect\citeauthoryear{Maslov}{2021}]{Maslov2021} Maslov A.~I., 2021, ATsir, 1648, 1. doi:10.24412/0236-2457-2022-1648-1-3
\bibitem[\protect\citeauthoryear{McKinney}{2010}]{pandas2010} McKinney W., 2010, scpy.soft. doi:10.25080/Majora-92bf1922-00a
\bibitem[\protect\citeauthoryear{Mies}{1974}]{Mies1974} Mies F.~H., 1974, ApJL, 191, L145. doi:10.1086/181572
\bibitem[\protect\citeauthoryear{Morbidelli et al.}{2000}]{Morbidelli2000} Morbidelli A., Chambers J., Lunine J.~I., Petit J.~M., Robert F., Valsecchi G.~B., Cyr K.~E., 2000, M\&PS, 35, 1309. doi:10.1111/j.1945-5100.2000.tb01518.x
\bibitem[\protect\citeauthoryear{Moreno et al.}{2025}]{Moreno2025} Moreno F., Goetz C., Aceituno F.~J., Casanova V., Sota A., Santos-Sanz P., 2025, MNRAS, 539, 949. doi:10.1093/mnras/staf552
\bibitem[\protect\citeauthoryear{Mugrauer et al.}{2024a}]{Mugrauer2024a} Mugrauer M., Michel K.-U., Pietsch L., Tschirschky A., 2024, ATel, 16887, 1
\bibitem[\protect\citeauthoryear{Mugrauer}{2024b}]{Mugrauer2024b} Mugrauer M., 2024, ATel, 16911, 1
\bibitem[\protect\citeauthoryear{Munaretto et al.}{2026}]{Munaretto2026} Munaretto G., Cambianica P., Cremonese G., Mura A., Ilyin I., Cusano F., Kuhn O., et al., 2026, Icar, 455, 117124. doi:10.1016/j.icarus.2026.117124
\bibitem[\protect\citeauthoryear{Mumma \& Charnley}{2011}]{Mumma2011} Mumma M.~J., Charnley S.~B., 2011, ARA\&A, 49, 471. doi:10.1146/annurev-astro-081309-130811
\bibitem[\protect\citeauthoryear{Sakai et al.}{2025}]{Sakai2025} Sakai N., Poshyachinda S., Sugiyama K., Rujopakarn W., Soonthornthum B., Leckngam A., Kramer B.~H., et al., 2025, PSJ, 6, 261. doi:10.3847/PSJ/ae0e19
\bibitem[\protect\citeauthoryear{Schleicher \& A'Hearn}{1988}]{Schleicher1988} Schleicher D.~G., A'Hearn M.~F., 1988, ApJ, 331, 1058. doi:10.1086/166622
\bibitem[\protect\citeauthoryear{Sekanina}{2024a}]{Sekanina2024a} Sekanina Z., 2024, arXiv, arXiv:2407.06166. doi:10.48550/arXiv.2407.06166
\bibitem[\protect\citeauthoryear{Sekanina}{2024b}]{Sekanina2024b} Sekanina Z., 2024, arXiv, arXiv:2407.11938. doi:10.48550/arXiv.2407.11938
\bibitem[\protect\citeauthoryear{Smith et al.}{2021}]{Smith2021} Smith A.~J., Anish Roshi D., Manoharan P., Vaddi S., Perera B.~B.~P., McGilvray A., 2021, PSJ, 2, 123. doi:10.3847/PSJ/abfec7
\bibitem[\protect\citeauthoryear{Tang et al.}{2021}]{Tang2021} Tang N., Li D., Yue N., Zuo P., Liu T., Luo G., Chen L., et al., 2021, ApJS, 252, 1. doi:10.3847/1538-4365/abca94
\bibitem[\protect\citeauthoryear{Tang et al.}{2024}]{Tang2024} Tang Y., Wang S., Lin Z., Yang X., Zhang X., Jia S., Wang S.~X., 2024, RNAAS, 8, 269. doi:10.3847/2515-5172/ad891a
\bibitem[\protect\citeauthoryear{The pandas development Team}{2025}]{pandas2025} The pandas development Team, 2025, zndo. doi:10.5281/zenodo.3509134
\bibitem[\protect\citeauthoryear{Vander Donckt et al.}{2026}]{VanderDonckt2026} Vander Donckt M., Jehin E., Aravind K., Adami C., Hmiddouch S., Manfroid J., Ganesh S., et al., 2026, A\&A, 705, A89. doi:10.1051/0004-6361/202556202
\bibitem[\protect\citeauthoryear{Virtanen et al.}{2020}]{scipy2020} Virtanen P., Gommers R., Oliphant T.~E., Haberland M., Reddy T., Cournapeau D., Burovski E., et al., 2020, NatMe, 17, 261. doi:10.1038/s41592-019-0686-2
\bibitem[\protect\citeauthoryear{Wang et al.}{2017}]{Wang2017} Wang Z., Chen X., Gao F., Zhang S., Zheng X.-W., Ip W.-H., Wang N., et al., 2017, AJ, 154, 249. doi:10.3847/1538-3881/aa97db
\bibitem[\protect\citeauthoryear{Winiarski, Waniak, \& Magdziarz}{1992}]{Winiarski1992} Winiarski M., Waniak W., Magdziarz P., 1992, EM\&P, 59, 229. doi:10.1007/BF00054055
\bibitem[\protect\citeauthoryear{Zhang et al.}{2023}]{Zhang2023} Zhang C.-P., Jiang P., Zhu M., Pan J., Cheng C., Liu H.-F., Zhu Y., et al., 2023, RAA, 23, 075016. doi:10.1088/1674-4527/acd58e
\bibitem[\protect\citeauthoryear{Zhao et al.}{2025}]{Zhao2025} Zhao R., Yang B., Kelley M.~S.~P., Protopapa S., Li A., Huang Y., Liu J., 2025, ApJ, 989, 146. doi:10.3847/1538-4357/adec8d

\end{thebibliography}

\bibliographystyle{aasjournal}



\end{document}